\documentclass{sig-alternate}

\usepackage{cite}
\usepackage{multirow}
\usepackage{algorithm}
\usepackage{algorithmic}
\usepackage{graphicx}
\usepackage{subfig}
\usepackage{balance}

\newcommand{\argmax}{\operatornamewithlimits{argmax}}

\newcommand{\esig}{expansion signature}
\newcommand{\esigcap}{Expansion Signature}
\newcommand{\greedy}{GreedyAPX}

\newdef{defn}{Definition}

\newtheorem{prop}{Proposition}

\begin{document}
\conferenceinfo{CIKM'10,} {October 26--30, 2010, Toronto, Ontario, Canada.}

\CopyrightYear{2010}

\crdata{978-1-4503-0099-5/10/10}

\clubpenalty=10000

\widowpenalty = 10000

\title{Expansion and Search in Networks}

\numberofauthors{2}

\author{
\alignauthor Arun S. Maiya\\
       \affaddr{Department of Computer Science}\\
       \affaddr{University of Illinois at Chicago}\\
       \affaddr{851 S. Morgan Street}\\
       \affaddr{Chicago, Illinois 60607}\\
       \email{amaiya@cs.uic.edu}
\alignauthor Tanya Y. Berger-Wolf\\
       \affaddr{Department of Computer Science}\\
       \affaddr{University of Illinois at Chicago}\\
       \affaddr{851 S. Morgan Street}\\
       \affaddr{Chicago, Illinois 60607}\\
       \email{tanyabw@cs.uic.edu}
}

\newcounter{copyrightbox}

\maketitle
\begin{abstract}
Borrowing from concepts in expander graphs, we study the expansion properties of real-world, complex networks (e.g. social networks, unstructured peer-to-peer or P2P networks) and the extent to which these properties can be exploited to understand and address the problem of decentralized search.  We first produce samples that concisely capture the overall expansion properties of an entire network, which we collectively refer to as the \emph{expansion signature}.  Using these signatures, we find a correspondence between the magnitude of maximum expansion and the extent to which a network can be efficiently searched.  We further find evidence that standard graph-theoretic measures, such as average path length, fail to fully explain the level of ``searchability'' or ease of information diffusion and dissemination in a network.  Finally, we demonstrate that this high expansion can be leveraged to facilitate decentralized search in networks and show that an expansion-based search strategy outperforms typical search methods.

\end{abstract}

\category{H.3.3}{Information Storage and Retrieval}{Information Search and Retrieval}[Retrieval Models]
\terms{Algorithms; Experimentation, Measurement} 
\keywords{expansion, decentralized search, P2P, peer-to-peer networks, social network analysis, complex networks, graph mining, expander graphs, focused web crawling, MANET}

\section{Introduction}
\label{sec:intro}
Motivated by the algorithmic problem of searching large graphs, we study the expansion properties of real-world networks and the extent to which these properties can be exploited to better understand and facilitate decentralized search.  Complex networks of linked entities arise across diverse domains, from sociology (e.g. social networks) to biology (e.g. neural networks, protein interactions) to technological and information systems (e.g. P2P networks, the Web, power grids).  Scientists from disparate fields, especially within the past decade, have attempted to characterize both the structure and function of these networked systems.  The standard approach here has been to measure topological features of the graph representing the network and correlate them with functional or dynamic aspects of the network (e.g. evolution of the network, the behavior of processes that occur over the network).  For instance, in their seminal paper on small-world networks, Watts and Strogatz \cite{Watts1998Collective} showed that many real-world networks simultaneously exhibit short average path lengths and relatively high degrees of clustering.  They further showed that these features can facilitate spreading processes across a network (e.g. the spread of a virus) \cite{Watts1998Collective}.  In this work, we study a feature that has received comparatively less attention in the study of real-world\footnote{There is a large body of work studying expansion in theoretical computer science and graph theory.  However, much of this work focuses on 1) synthetic graphs that do not normally arise in the real-world such as $d$-regular graphs and 2) the \emph{minimum} (not maximum) expansion in these graphs \cite{Hoory2006Expander}.} networks: \emph{expansion}.

\subsection{Expansion}
Given a network $G=(V,E)$ where $V$ is the set of nodes and $E$ is the set of links among the nodes, the \emph{expansion} of a set of nodes $S \subset V$ is a function of the number of nodes in $V-S$ to which $S$ is connected (see Section \ref{sec:preliminaries} for more precise definitions).  That is, if $N(S)$ is the set of nodes to which $S$ is connected, then the \emph{expansion} of $S$ is $\frac{|N(S)|}{|S|}$.  Informally, an expander graph is a graph in which any subset of nodes has good expansion (i.e. has many neighbors) \cite{Hoory2006Expander}.  For instance, a network is said to be a $\gamma$-expander if every $S \subset V$ has an expansion of \emph{at least} $\gamma$, where $|S| < \frac{|V|}{2}$ \cite{Hoory2006Expander}.  Thus, the classical definition of an expander graph focuses on samples with the \emph{minimum} expansion (as $\gamma$ is a minimum if every sample $S$ has an expansion of at least $\gamma$). Expander graphs have been shown to have many applications from constructing error correcting codes to routing calls in telephone switching networks \cite{Hoory2006Expander}.  Although these classical expander graphs are well-studied, there has been surprisingly less attention paid to studying the sample in a network with the \emph{maximum} expansion, and it is this in which we are most interested.  Our overall aim in this work is to investigate the extent to which a vertex set with high expansion can be leveraged to both understand and facilitate efficient decentralized search in networks.  We now describe the problem of decentralized search.

\subsection{Search}
A central algorithmic problem in the study of complex networks is how to efficiently search them in a decentralized manner.  This scenario arises in many practical applications from querying peer-to-peer file sharing networks to focused Web crawling \cite{Kleinberg2006Complex}.  Starting from some initial source node, we must locate, access, or route a message to some other target node in the network.   Without full knowledge of the global network topology, we are unable to simply compute the shortest path or access the target node directly.  Thus, we must hop from node to node until the target node is found.  Decentralized search, then, is related to information diffusion or dissemination, as an efficient search will involve efficiently disseminating a query message to large portions of the network.  In this work, we study the effect of network expansion on decentralized search.  If, using only local information, nodes are visited in such a way that their overall distance is close to many other nodes (i.e. the set of visited nodes has high expansion), then the efficiency of search might be improved. Moreover, the magnitude of expansion in a network may shed light on the extent to which a network can be efficiently searched by \emph{any} search algorithm.  These are precisely the questions we investigate here. 

\subsection{Contributions and Summary of Findings}
\label{sec:intro.contributions}
   
We rigorously investigate decentralized search across a span of networks much wider and more diverse than previously studied in work on searching graphs. Our main contributions include the following:

\begin{itemize}
 \item Borrowing from concepts in expander graphs, we introduce the concept of \emph{\esig s}, which concisely captures the overall expansion properties of a network.  We find that, in many networks, relatively small samples of nodes can exhibit significantly high expansion. However, we also find that there are a few networks for which small samples with high expansion may not exist. 
\item We propose an expansion-based, decentralized search strategy, which explicitly tries to locate these samples with high expansion in order to quickly discover the most nodes in a search.  We evaluate a number of search algorithms such as degree-based searches, breadth-first searches, and random walks.  We show that an expansion-based search strategy generally outperforms others.
\item We demonstrate that \emph{\esig s} correctly infer the extent to which a network can be efficiently searched (which we refer to as ``searchability'').  Moreover, we show that it is the maximum expansion in a network, rather than the minimum expansion, that contributes most to searchability and information dissemination in a network.  At the same time, we find that standard graph-theoretic measures, such as average path length, \emph{fail} to fully explain the extent to which a network is easily searched.
\end{itemize}

The last point is our most significant finding.  Existing works have mostly studied the effect of \emph{minimum} expansion on the ease of dissemination in a network (e.g. \cite{Hoory2006Expander,Chierichetti2010Rumour}).  Other works have focused on the effect of standard graph properties such as average path length on searchability and ease of dissemination (e.g. \cite{Hui2006Smallworld}).  For the first time, we show that it is the \emph{maximum} expansion (rather than minimum expansion) that most affects efficient searchability.  Our results, then, offer a more comprehensive picture of decentralized search and information diffusion in networks than has previously been appreciated.

\section{Background and Related Work}
\label{sec:relatedwork}

Interestingly, one of the first experiments on decentralized search in networks was the famous chain-letter study by the social psychologist Stanley Milgram \cite{Milgram1967Small}.  In this experiment, participants were given the name, address, and occupation of an unknown target person and told to forward a chain letter to this person by passing the letter on to a single acquaintance meeting two main conditions:  1) the acquaintance must be someone with whom the individual knew on a first-name basis and 2) the acquaintance chosen should be the one perceived as closest to the target \cite{Milgram1967Small}.  This study not only provided some evidence for short paths in social networks\footnote{As noted in \cite{Kleinfeld2002Could}, a number of issues exist in Milgram's results.  For instance, many chain letters failed to ever reach the target.  Nevertheless, the conclusion that short path lengths exist in social networks is generally accepted today and has been verified in many networked data \cite{Kleinberg2006Complex}.} (the median path length between sources and targets was $6$ among letters that reached their destination), but also showed that individuals were able to collectively discover these short paths \emph{without} full knowledge of the network \cite{Kleinberg2000SmallWorld}.  Kleinberg \cite{Kleinberg2000SmallWorld} later modeled this problem algorithmically using a 2-dimensional grid with probabilistically-added long-range connections and shed light on the precise conditions that these short paths were discoverable using a decentralized search algorithm.  

In both the works of Milgram \cite{Milgram1967Small} and Kleinberg \cite{Kleinberg2000SmallWorld}, although nodes had no knowledge of global network connectivity, there was, in fact, \emph{external} knowledge that aided searches.  In the Milgram experiment, for instance, individuals were instructed to forward letters to the local acquaintance that was perceived as being closest to the target - as measured by geographic or occupational similarity in many cases.  Thus, external knowledge of geographical distance and occupational similarity was employed as an aid in the search heuristic.  In other words, there was knowledge (or, at least, an assumption) that individuals closer geographically or more similar occupationally are more likely to know each other.  Liben-Nowell et al. \cite{LibenNowell2005Geographic}, in fact, showed some evidence for the geographical basis of online friendships in the LiveJournal social network.  In Kleinberg's model also, it is assumed that message holders have access to external knowledge and know the local contacts of \emph{all} nodes in the network (i.e. nodes are aware of the Manhattan distances between all nodes in the underlying grid structure and use this information as a forwarding heuristic).  Boguna et al. \cite{Boguna2008Navigability} have recently modeled this external information as a hidden metric space.  

Unfortunately, in many real-world scenarios, such as unstructured peer-to-peer file sharing networks, these types of external information and similarity-based heuristics are unavailable as search aids, and the problem of decentralized search becomes even more challenging.  Typical approaches here resort to variations on flooding the network (which can be unscalable), random walks (which may be less effective in finding information), or imposing structure on the network to improve searchability (which requires additional overhead) \cite{Tsoumakos2006Analysis}.  For a review of decentralized search both in the contexts of complex networks and specifically P2P, one may refer to \cite{Kleinberg2006Complex,Mitra2009Technological,Tsoumakos2006Analysis}.  Our focus in this work is to investigate efficient search on networks with \emph{arbitrary} structure in which similarity-based heuristics (e.g. geographic distance) are unavailable.  For the first time, we investigate the relationship between \emph{expansion} and the extent to which a network is efficiently searchable.  We further show that a search strategy based on expansion generally outperforms typical existing approaches such as random walks and flooding-based techniques.  We begin a discussion of our work with some preliminaries.

\section{Preliminaries}
\label{sec:preliminaries}

\subsection{Notations and Definitions}
\label{sec:preliminaries.definitions}

We now briefly describe some notations and definitions used throughout this paper.

\begin{defn}
\label{defn:network}
$G=(V,\,E)$ is an undirected \emph{network} or \emph{graph} where $V$ is a set of vertices (or nodes) and $E \subseteq V \times V$ is a set of edges (or links between the nodes).  We will use the terms \emph{network} and \emph{graph} interchangeably.  
\end{defn}

\begin{defn}
\label{defn:sample}
A \emph{sample} $S$ is a subset of vertices, $S \subset V$.
\end{defn}

\begin{defn}
\label{defn:neighborhood}
$N(S)$ is the \emph{neighborhood} of $S$ if $N(S)=\{w \in V-S: \; \exists v \in S \; s.t. \; (v,\,w) \in E\}$.  The \emph{neighborhood} may also be referred to as the \emph{frontier} of a sample $S$. 
\end{defn}

\begin{defn}
\label{defn:expansion}
The \emph{expansion} of a sample\footnote{The \emph{expansion} of an entire graph is typically taken to mean $\min_{S \subset V}\frac{|N(S)|}{|S|}$ \cite{Hoory2006Expander}.} $S$ is: $$\frac{|N(S)|}{|S|}$$  
\end{defn}

\begin{defn}
\label{defn:expanderset}
The \emph{maximum expander set} of size $k$ is a sample $S$ of size $k$ with the maximal expansion:  $$\argmax_{S:\,|S|=k} \frac{|N(S)|}{|S|}$$
\end{defn}

\begin{defn}
\label{defn:expansionquality}
The \emph{expansion quality} of a sample $S$ is the normalized\footnote{Alternatively, one can normalize \emph{expansion} as $\frac{|N(S) \cup S|}{|V|}$.} \emph{expansion}:  
$$ \frac{|N(S)|}{|S|} \div \frac{|V-S|}{|S|} = \frac{|N(S)|}{|V-S|}$$.
\end{defn}
Notice that, given a sample $S$, the maximum possible expansion on \emph{any} network of $|V|$ nodes is:  $\frac{|V-S|}{|S|}$.
The \emph{expansion quality} $\frac{|N(S)|}{|V-S|}$, then, captures the extent to which a sample achieves this maximum possible expansion.  A score of $1$ indicates that the sample ``touches'' or is one hop away from every other node in the network.
 
\subsection{Datasets}
\label{sec:preliminaries.datasets}

We study expansion and search in a total of ten different networks:  two random graph models, a neural network, a power grid, a co-authorship network, an email network, a citation network, a P2P file-sharing network, and two online social networks.  It should be noted that not all of these networks may require efficient decentralized search (e.g. a co-authorship network, the neural network of a worm).  Nevertheless, these datasets represent a rich set of diverse networks from different domains.  This allows us to more comprehensively study network expansion and thoroughly assess the performance of decentralized search strategies in the face of varying network topologies.  Table \ref{tab:datasets} shows characteristics of each network.  We now describe each dataset.

\noindent
\textbf{Erdos-Renyi Model.}  One of the first random graph models proposed was that of Erdos and Renyi \cite{Erdos1959Random}.  The Erdos-Renyi $G(n,p)$ model produces a random graph of $n$ nodes with each of the ${n \choose 2}$ possible edges existing with probability $p$.  Erdos-Renyi graphs exhibit the short average path lengths found in many real-world networks, but lack the high clustering and skewed (or heavy-tailed) degree distributions found in reality.  

\noindent
\textbf{Barabasi-Albert Model.}   The Barabasi-Albert model follows a more, realistic generative process than previous models: the preferential attachment model \cite{Barabasi1999Emergence}.  A graph of $n$ nodes is grown in a sequential fashion.  Each subsequent node of $m$ edges is preferentially attached to previously added nodes with  high degree (where the ``degree'' of a node is the number of neighbors).  Graphs generated by this model exhibit skewed, power law degree distributions and short average path lengths, but lack the high clustering found in real networks.  (Skewed degree distributions are ones in which there are many nodes with low connectivity and a few nodes with high connectivity that act as  hubs.  A power law distribution is one such example.)

\noindent
\textbf{C. elegans Neural Network} is the neural network of the C. elegans worm \cite{Watts1998Collective}.

\noindent
\textbf{Power Grid.}  This technological network represents the power grid of the western United States \cite{Watts1998Collective}.

\noindent
\textbf{CondMat.}  This is a co-authorship network of scientists publishing in Arxiv Cond-Mat (i.e. the Condensed Matter Physics category) from the e-print archive, arxiv.org \cite{Leskovec2005Graphs}.

\noindent
\textbf{Enron Emails} is the network comprised of email communications among Enron employees \cite{Klimt2004Enron}.

\noindent
\textbf{HEPPh} is a citation network between papers in Arxiv HEP-Ph (high energy physics phenomenology) from the e-print archive, arxiv.org \cite{Leskovec2005Graphs,Gehrke2003Overview}.

\noindent
\textbf{Gnutella.} This network is an August 31st, 2002 snapshot of the Gnutella peer-to-peer file-sharing network.  Nodes represent hosts and edges represent connections among the hosts \cite{Ripeanu2002Mapping}.

\noindent
\textbf{Epinions} is a trust-based online social network of the consumer review site, Epinions.com \cite{Richardson2003Trust}.

\noindent
\textbf{Slashdot} is an online social network of the technology news site, Slashdot.com \cite{Leskovec2008Statistical}.

\begin{table} [th]
\centering
\footnotesize
\begin{tabular}{l|c|c|c|c|c} \hline \hline
 Random Graphs    &  N     & D                        & PL   & CC       & AD\\ \hline
 Erdos-Renyi      & 10,000   & 0.0005                 & 4.2  & 0.0005   & 6.0 \\
 Barabasi-Albert  & 10,000   & 0.0005                 & 3.0  & 0.006    & 6.0   \\ \hline \hline
Real-World        &  N     & D                        & PL   & CC       & AD \\ \hline
 C. elegans       & 297   & 0.05                      & 2.5  & 0.3      & 14.5 \\
 Power Grid       & 4941   & 0.0005                   & 19   & 0.11     & 2.7 \\
 CondMat          & 21,363 & 0.0004                   & 5.4  & 0.70     & 8.5 \\
 Enron            & 33,696 & 0.0003                   & 4.0    & 0.71     & 10.7 \\
 HEPPh            & 34,401 & 0.0007                   & 4.3    & 0.30     & 24.5 \\
 Gnutella         & 62,561 & 0.00008                  & 5.9    & 0.01     & 4.7  \\
 Epinions         & 75,877  & 0.0001                  & 4.3    & 0.26      & 10.7 \\
 Slashdot         & 82,168  & 0.0001                  & 4.1    & 0.10      & 12.2 \\ \hline \hline
\end{tabular}
\caption{Network Properties.  \textbf{Key:}  \emph {N= \# of nodes, D= density, PL = characteristic path length, CC = clustering coefficient, AD = average degree.}}
\label{tab:datasets}
\end{table}

\section{Expansion Signatures}
\label{sec:expansion}

In this section we introduce the concept of \emph{\esig s}, which concisely captures the expansion properties of a network at different size scales.  Intuitively, the \emph{\esig} plots the maximum (and minimum) \emph{expansion qualities} of samples at increasing sample sizes.  As discussed, we are mostly interested in samples with the \emph{maximum} expansion, but we include the \emph{minimum} expansion for completeness.  As we will see later, \emph{\esig s} reveal a number of interesting aspects of networks.  But first, we address how precisely to compute the \emph{\esig} of a network.

\subsection{Problem Formalization}

To construct the \emph{\esig}, we must seek out the sample $S$ of size $k$ with the maximal (and minimal) expansion for progressively increasing values of $k$.  In Definition \ref{defn:expanderset}, we defined the \emph{maximum expander set} as the sample of size $k$ with the maximum expansion.  We now formally define the problem of finding this sample:

\begin{defn}
(\emph{Maximum Expansion Problem}) Given a graph $G=(V,E)$ and a sample size $k < |V|$, the {\sc Maximum Expansion} problem (\textbf{MEP)} is to find a sample $S \subset V$  of size $k$ with the maximum expansion, $\frac{|N(S)|}{|S|}$.  That is, find: $\argmax_{S:\,|S|=k} \frac{|N(S)|}{|S|}$.
\end{defn}

The hardness of various problems related to expansion is well-studied (e.g. \cite{Hoory2006Expander}).  For instance, determining that a graph is a $\gamma$-expander (where every $S \subset V$ has expansion of at least $\gamma$ and $|S| < \frac{|V|}{2}$) is known to be co-NP-complete \cite{Hoory2006Expander}.  The {\sc Dominating Set} problem \cite{1997Approximation}, known to be NP-hard, is to find the smallest sample $S$ such that $N(S) = V-S$.  {\sc Maximum Expansion} is clearly a generalization of {\sc Dominating Set} and is, thus, also NP-hard.  There also exist reductions to and from the {\sc Maximum Coverage} problem (defined in \cite{1997Approximation} and also below).  Proposition \ref{prop:maxexpnphard}, for instance, shows NP-hardness by reduction from {\sc Maximum Coverage}.

\begin{prop}
\label{prop:maxexpnphard}
The {\sc Maximum Expansion} problem is NP-hard.
\end{prop}
\begin{proof}
 We show a reduction from the {\sc Maximum Coverage} problem, which is known to be NP-hard \cite{1997Approximation}.  In {\sc Maximum Coverage}, given a set $\mathcal{U}$ of $n$ elements, a collection $\mathcal{F}=\{C_{i} \mid i \in I\}$ of $|I|$ subsets of $\mathcal{U}$ where $\bigcup_{i}C_{i}=\mathcal{U}$, and an integer $k < |I|$, the goal is to find $k$ subsets of $\mathcal{F}$ such that their union has the maximum cardinality.  To construct a graph $G$ for the {\sc Maximum Expansion} instance, for each $i\in I$ we create a node $i$.  For each element $u$ in $\mathcal{U}$, we create a node $u$.  Thus, $V=I \cup \mathcal{U}$ is the vertex set of $G$.  To create the edge set $E$ of $G$, an edge $\{i,j\}$ is created for each pair $i,j \in I$ (forming a clique among nodes in $I$).  In addition, for each $i \in I$ and $u \in C_i$, an edge $\{i,u\}$ is created (forming an independent set among nodes in $\mathcal{U}$).  

If $C=\{C_{i} \mid i \in S\}$ is a feasible solution to the {\sc Maximum Coverage} instance for some subset $S \subset I$ where $|S|=|C|=k$, then $S$ is a sample of nodes in $G$ with maximum expansion where $|S| = k$.   By construction, each set $C_i \in C$ (where $i \in S$) is represented by a node $i \in I$ from $G$ that is both connected to every other node in $I$ and connected to the nodes in $\mathcal{U}$ that represent elements {\em contained} by $C_i$. Thus, $S$ is a $k$-size sample with the largest neighborhood in $G$.  Conversely, let $S \subset V$ be a sample in $G$ with the maximum expansion.  Note that, if $S \cap \mathcal{U} \neq \emptyset$, then a new sample $S^\prime$, with $\frac{|N(S^\prime)|}{|S^\prime|} \geq \frac{|N(S)|}{|S|}$, can be constructed by replacing each node $v \in S \cap \mathcal{U}$  with one of $v$'s neighbors $w \in I$.  Thus, $C=\{C_{i} \mid i \in S^\prime\}$ is a feasible solution to {\sc Maximum Coverage}, since each node in $S^\prime$ represents a set in $\mathcal{F}$.
   
\end{proof}

Given the hardness of expansion-related problems, one typically resorts to spectral analysis, as the spectrum of a graph can be computed in polynomial time \cite{Hoory2006Expander}.  A key difference in our work, however, is that we are not only interested in the magnitude of expansion, but the \emph{identity} of the sample producing it.  Moreover, we are most interested in the \emph{maximum} expansion (as opposed to the minimum expansion, which is normally the focus in theoretical work on expander graphs).  Our ultimate objective is to access these high expansion nodes during the course of a decentralized search to understand and facilitate search performance.\footnote{For directed networks that are very weakly connected, a sample with high maximum expansion (based on out-degree) may exist, but the nodes in the sample itself may not be reachable from substantial portions of the network.  Samples such as this may shed little light on searchability.  One possible approach to address these scenarios is to compute \esig s using the \emph{expected} maximum expansion of \emph{connected} samples.  In the present work, however, for simplicity and brevity, we treat all links as bidirectional (or undirected).}  Spectral analysis may be less useful here. Thus, we approximate expansion using a simple greedy algorithm ({\sc \greedy}).  At each iteration, we greedily select the node that maximizes (or minimizes) the expansion of the currently constructed sample, as shown in Algorithm \ref{alg:greedy}.  We now show that this simple greedy algorithm yields a $(1-1/e)$-approximation guarantee for the {\sc Maximum Expansion} problem. 

\begin{prop}
\label{prop:maxexpapx}
{\sc GreedyAPX} approximates {\sc Maximum Expansion} within a ratio of at least $1 - 1/e \approxeq  0.632$.
\end{prop}
\begin{proof}
The structure of this proof follows that of the well-known proof for the {\sc Maximum Coverage} greedy approximation (see \cite{Hochbaum1998Analysis,Feige1998Threshold}, for instance).  Let $S_{opt}$ be the optimal sample of size $k$ and $N$ be the set of nodes covered by $S_{opt}$ (where ``covered'' is taken to mean $N(S_{opt}) \cup S_{opt}$).  Let $N_{i}$ be the set of \emph{new} nodes covered by the $i^{th}$ iteration of {\sc GreedyAPX}.  Since $N$ can be covered by a sample of size $k$, by the pigeonhole principle:
$$ |N_{i}| \geq \frac{|N| - \sum_{j=1}^{i-1} |N_{j}|}{k}$$
Then, $\sum_{j=1}^{i} |N_{j}| \geq |N| - |N|(1 - \frac{1}{k})^i$ and 
$$\sum_{i=1}^{k}|N_{i}| \geq |N| - |N|(1 - \frac{1}{k})^k \geq |N|(1-\frac{1}{e}).$$
\end{proof}

\begin{algorithm}[tb]
   \caption{{\sc \greedy}}
   \label{alg:greedy}
\begin{algorithmic}[1]
   \STATE {\bfseries Input:} \\\quad Graph $G=(V, E)$ \\ \quad $k$, the sample size. 
   \STATE $S = \emptyset$ \qquad \qquad \qquad // initialize sample to empty set
   \STATE $v = \argmax_w |N(\{w\})|$ \quad 
   \STATE $S = S \cup \{v\}$
   \WHILE{$|S| \leq k$}
   \STATE Select new node $v \in V-S$ that \\ maximizes (or minimizes): \\ \qquad \qquad $|N(\{v\})-(N(S) \cup S)|$
   \STATE $S = S \cup \{v\}$
   \ENDWHILE
\end{algorithmic}
\end{algorithm}

It should be noted that, during preliminary testing, we also experimented with using simulated annealing for finding the sample with maximum expansion, but {\sc \greedy} was shown to be superior.  We now use {\sc \greedy} to construct \emph{\esig s} for both synthetic random graphs and real-world networks.  We discuss each separately.

\subsection{Signatures for Random Graphs}

We examine the \emph{\esig s} of two well-known random graph models:  Erdos-Renyi (ER) graphs \cite{Erdos1959Random} and the Barabasi-Albert preferential attachment model (BA) \cite{Barabasi1999Emergence}.   The ER and BA models produce graphs with very different degree distributions.  Whereas the BA model produces graphs with highly skewed, heavy-tailed degree distributions that follow the power law \cite{Barabasi1999Emergence}, the ER model produces graphs following a Poisson degree distribution \cite{Erdos1959Random}. It is clear that the BA model exhibits a higher and more rapidly increasing maximum expansion (which is a result of the highly connected hubs in its skewed degree distribution).  However, the ER model exhibits a relatively higher \emph{minimum} expansion.  In fact, random $d$-regular graphs, where every node has the same degree $d$, also have ``good'' minimum expansion with high probability (where \emph{every} $S \subset V$ will have high expansion) \cite{Hoory2006Expander}.  In Section \ref{sec:evaluation}, we will examine whether it is the maximum or minimum expansion that most affects searchability.

\begin{figure}[ht]
  \centering
  \subfloat[Erdos-Renyi] {\label{fig:esig.er}\includegraphics[width=0.2\textwidth]{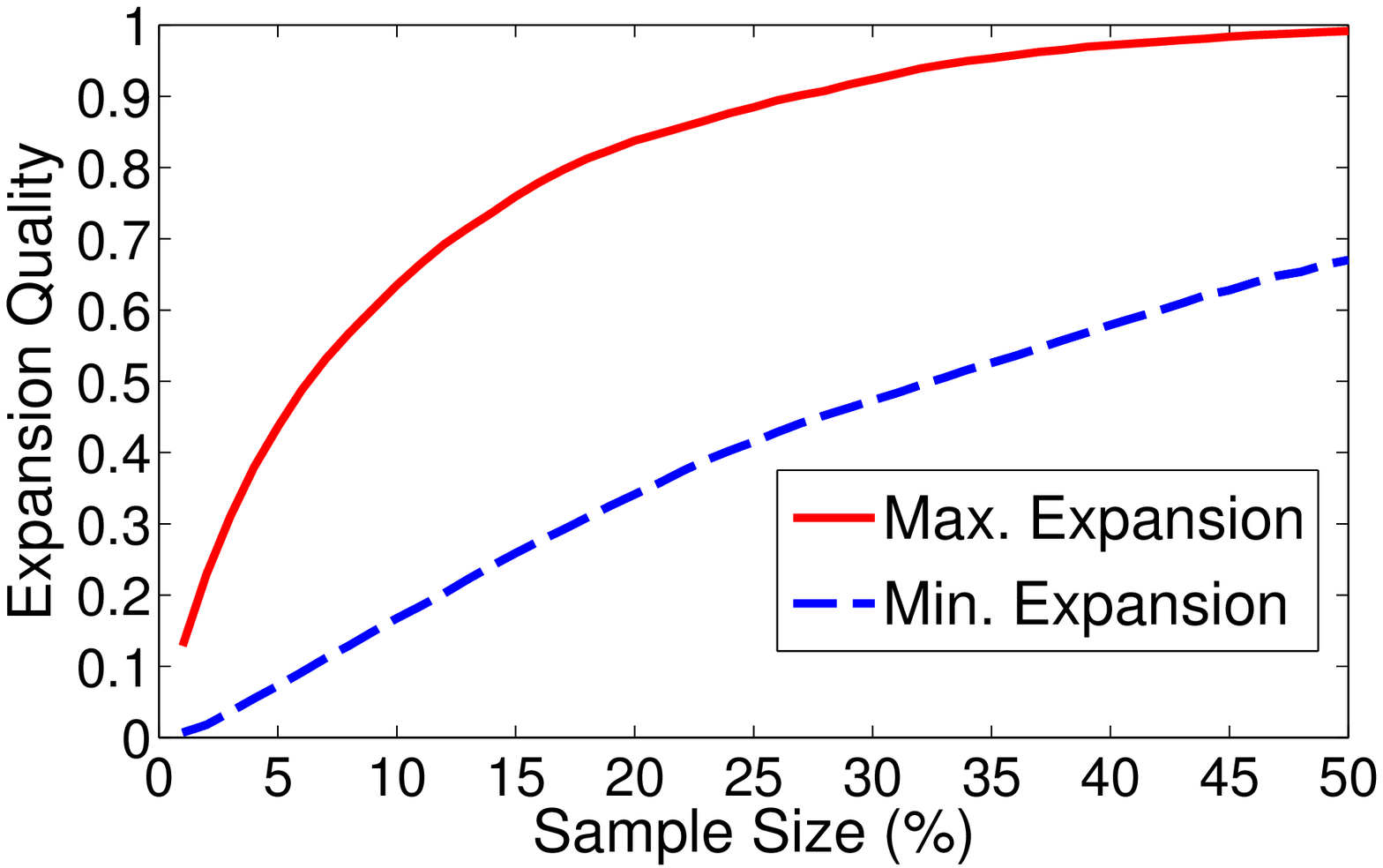}} \vspace{.05cm}
  \subfloat[Barabasi-Albert]{\label{fig:esig.ba}\includegraphics[width=0.2\textwidth]{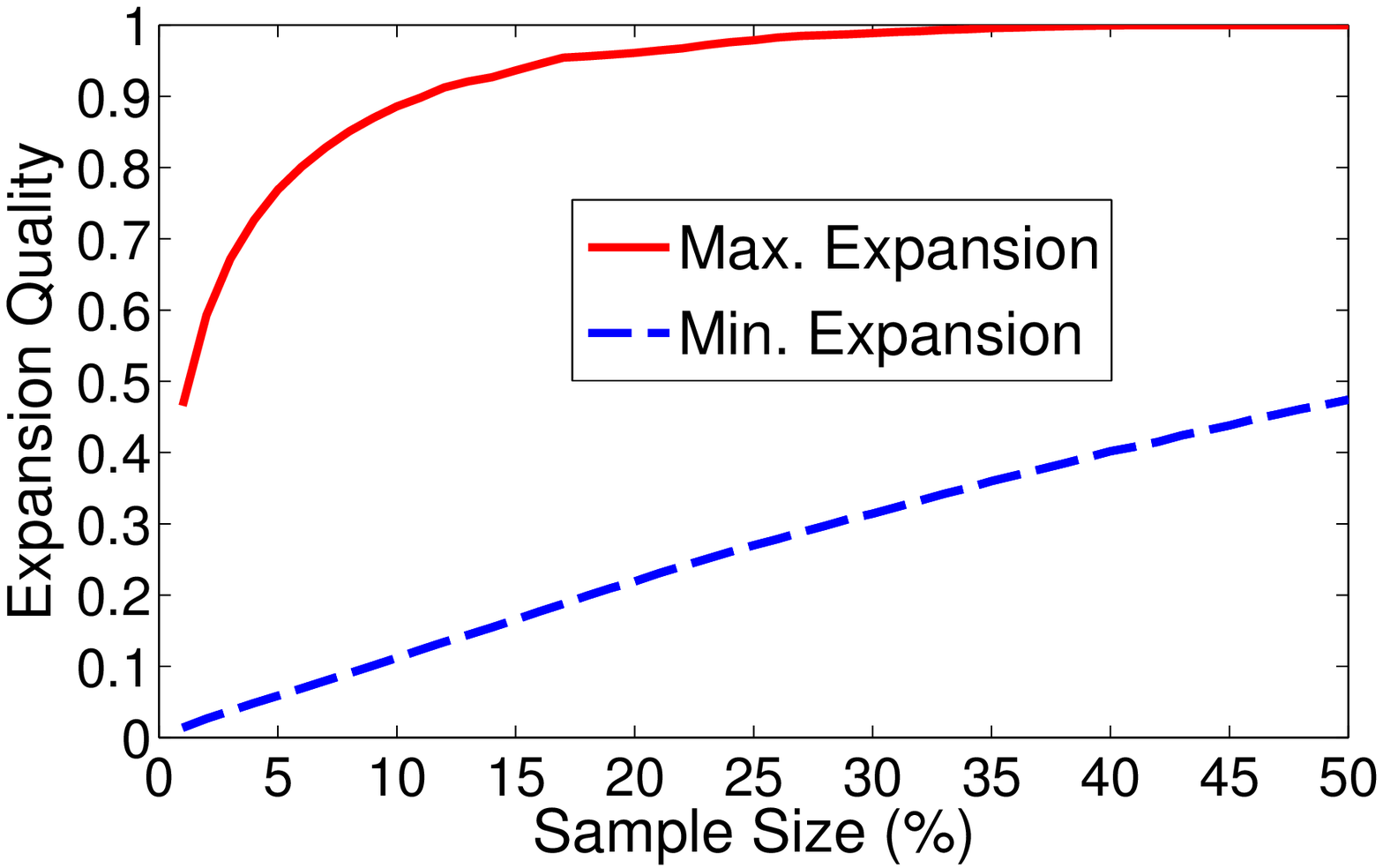}}
\caption{\emph{\esigcap s} for ER and BA models.}
  \label{fig:esig.randomgraphs}
\end{figure}

\subsection{Signatures for Real-World Networks}

\begin{figure}[ht]
  \centering
  \subfloat[C. elegans] {\label{fig:esig.celegans}\includegraphics[width=0.2\textwidth]{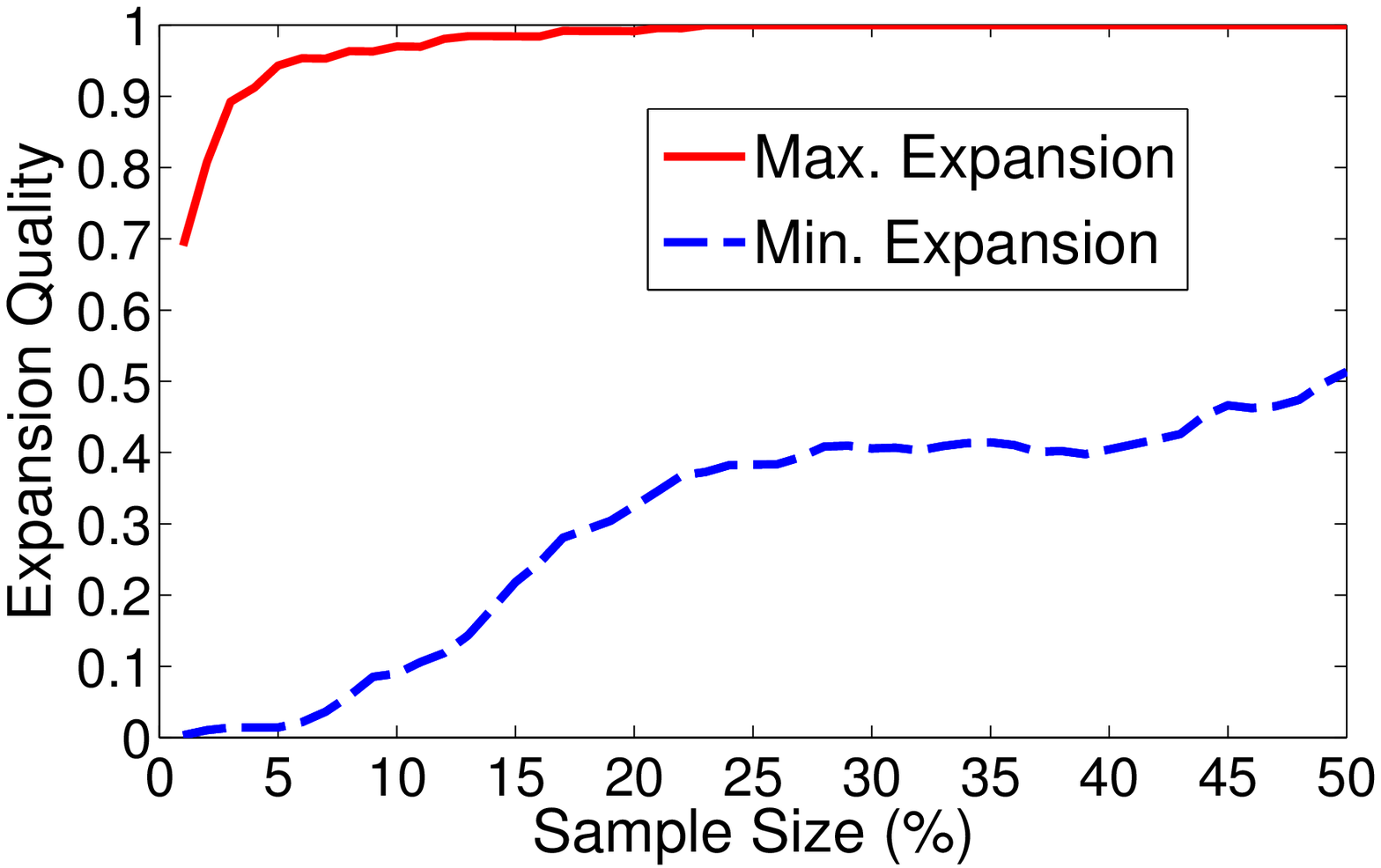}} \vspace{.05cm}
  \subfloat[Power Grid]{\label{fig:esig.power}\includegraphics[width=0.2\textwidth]{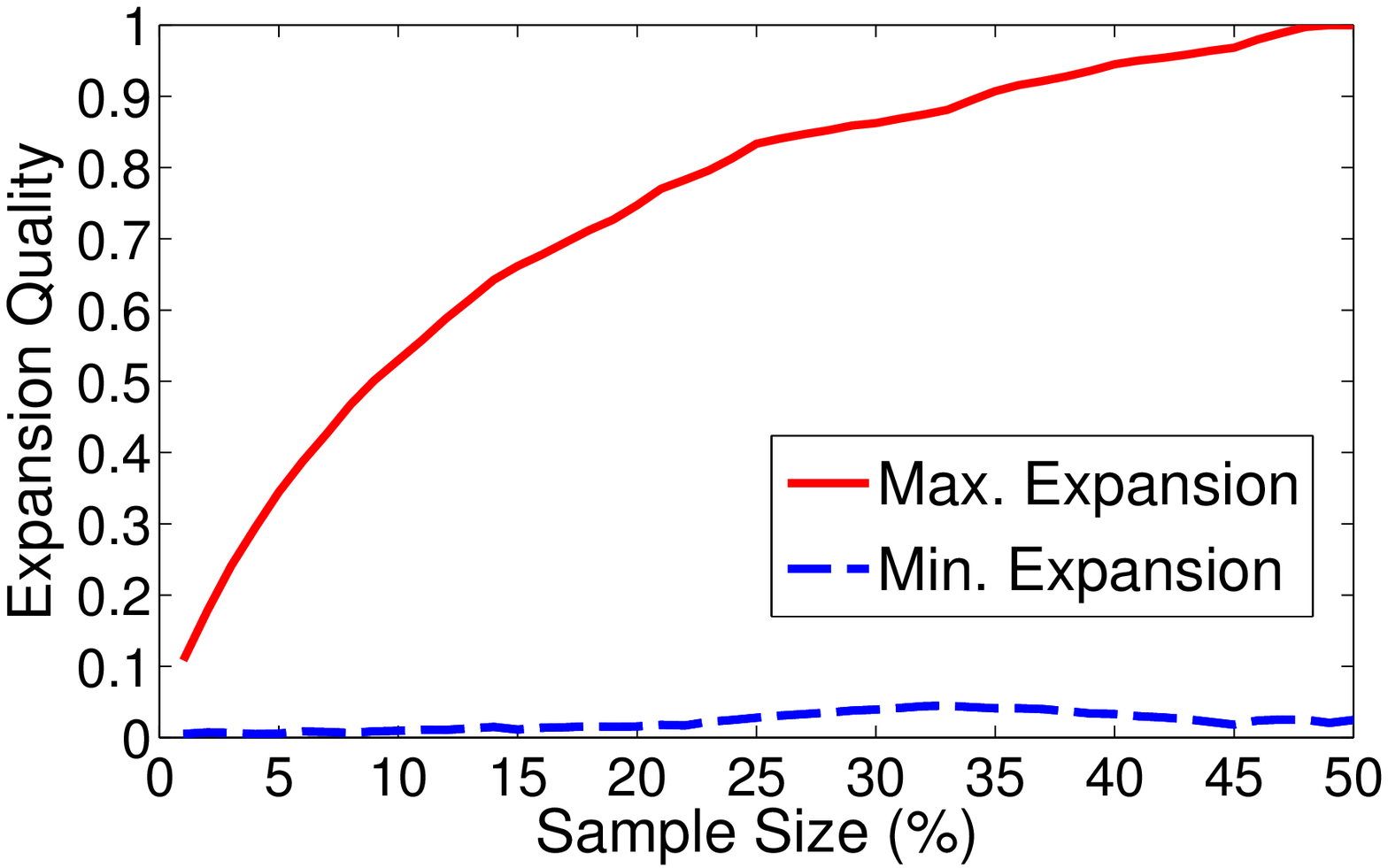}} \\ \vskip -0.01in 
  \subfloat[CondMat] {\label{fig:esig.condmat}\includegraphics[width=0.2\textwidth]{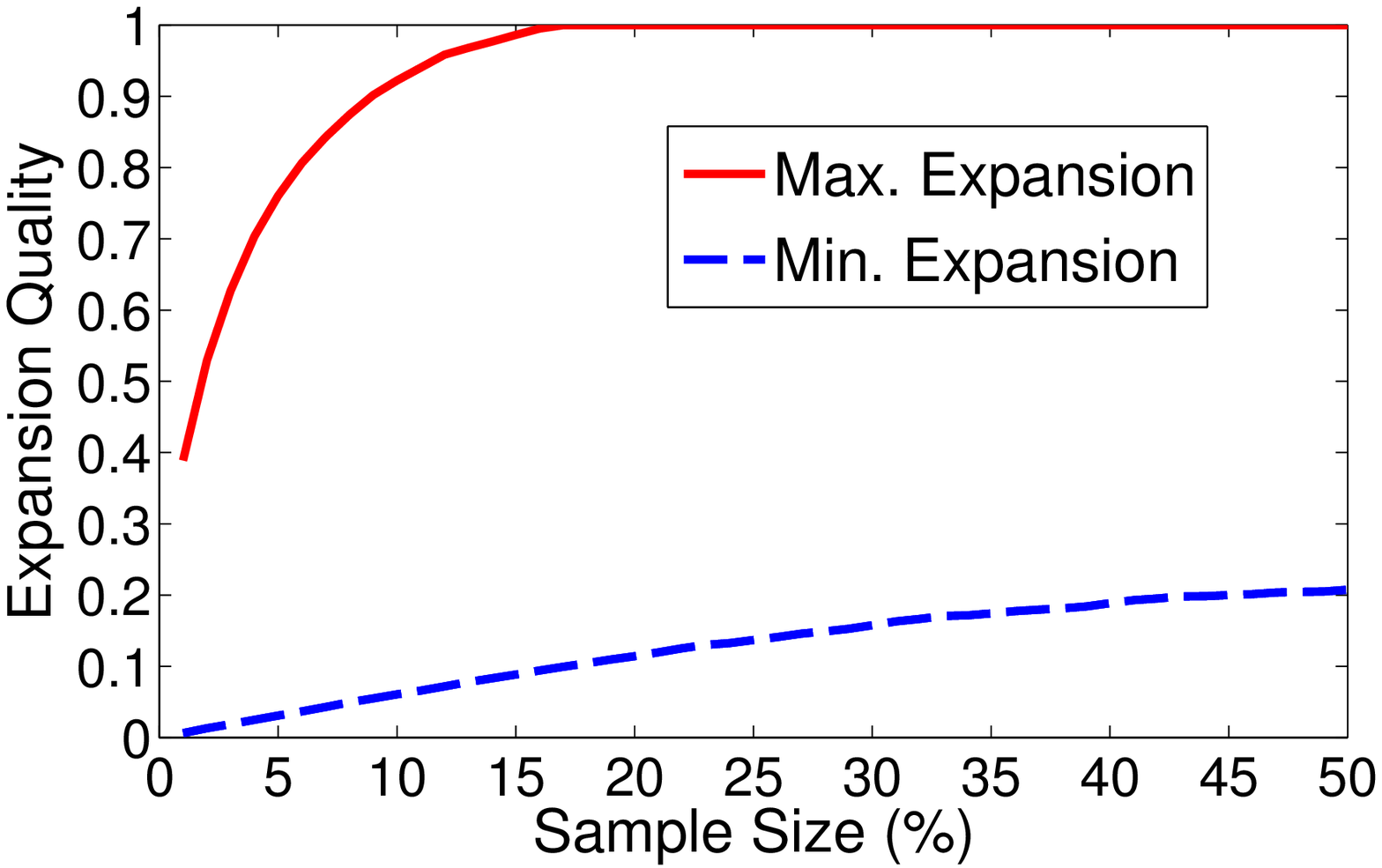}} \vspace{.05cm}
  \subfloat[Enron] {\label{fig:esig.enron}\includegraphics[width=0.2\textwidth]{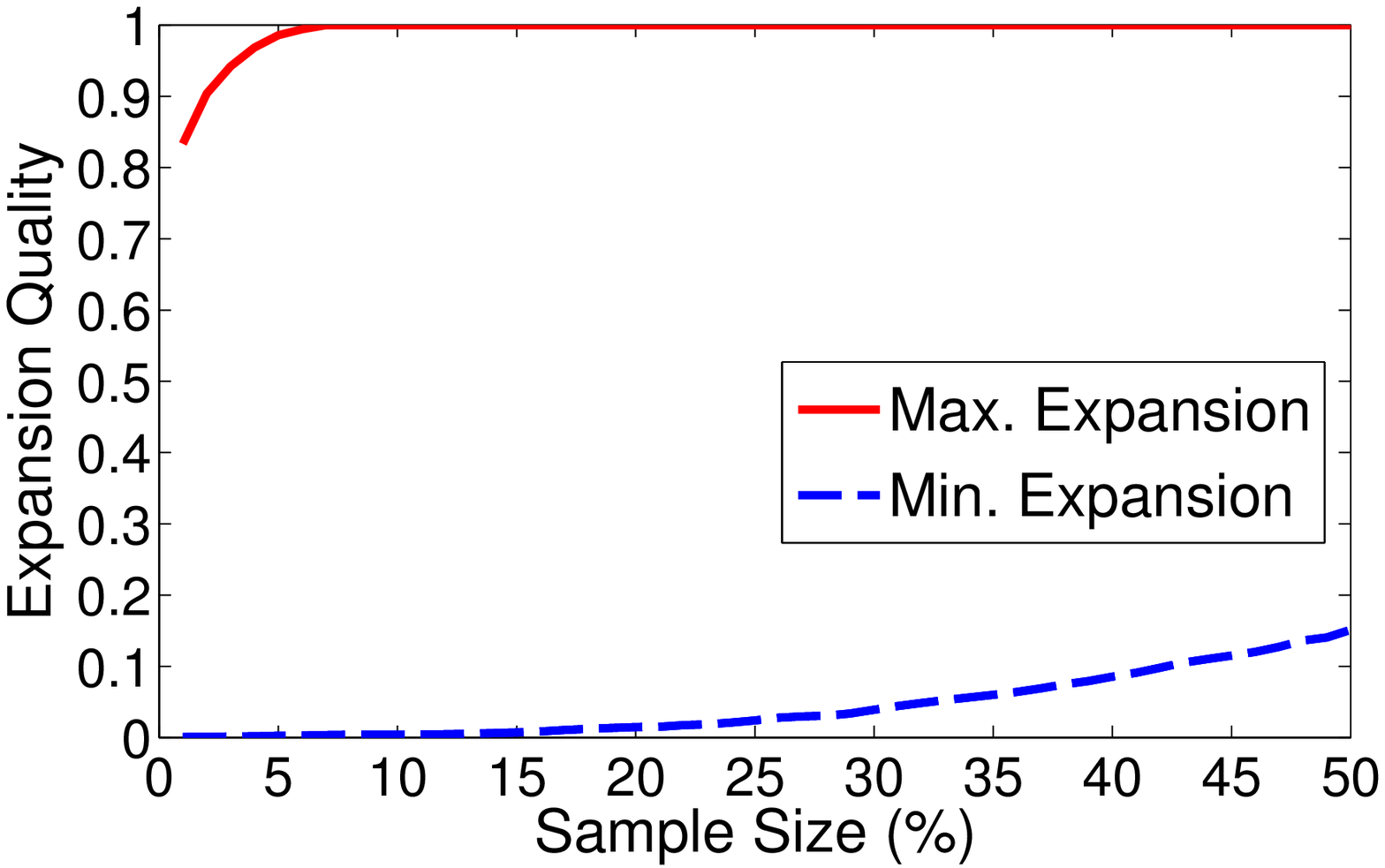}} \\ \vskip -0.01in
  \subfloat[HEPPh]{\label{fig:esig.hepph}\includegraphics[width=0.2\textwidth]{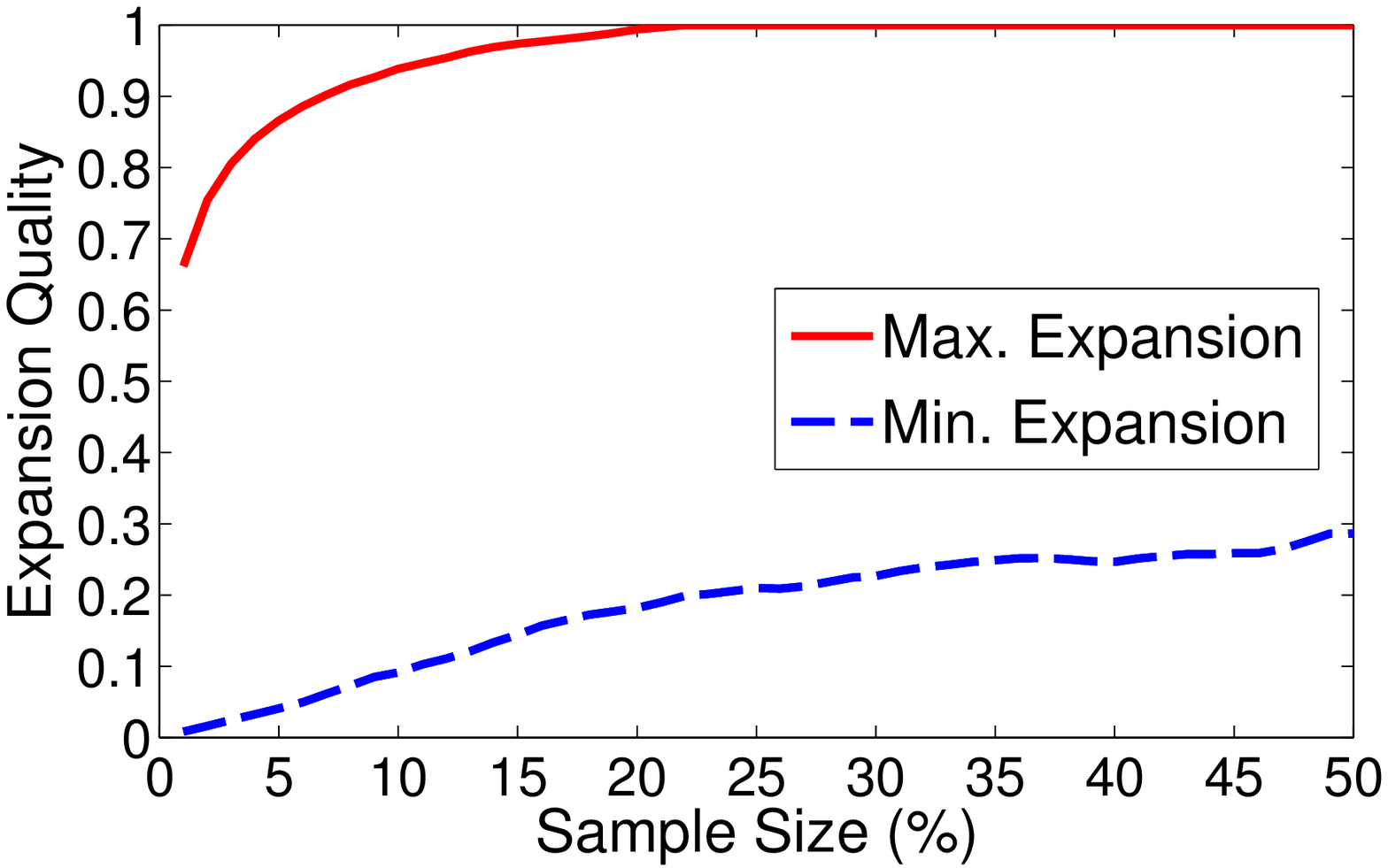}}\vspace{.05cm}
  \subfloat[Gnutella]{\label{fig:esig.gnutella}\includegraphics[width=0.2\textwidth]{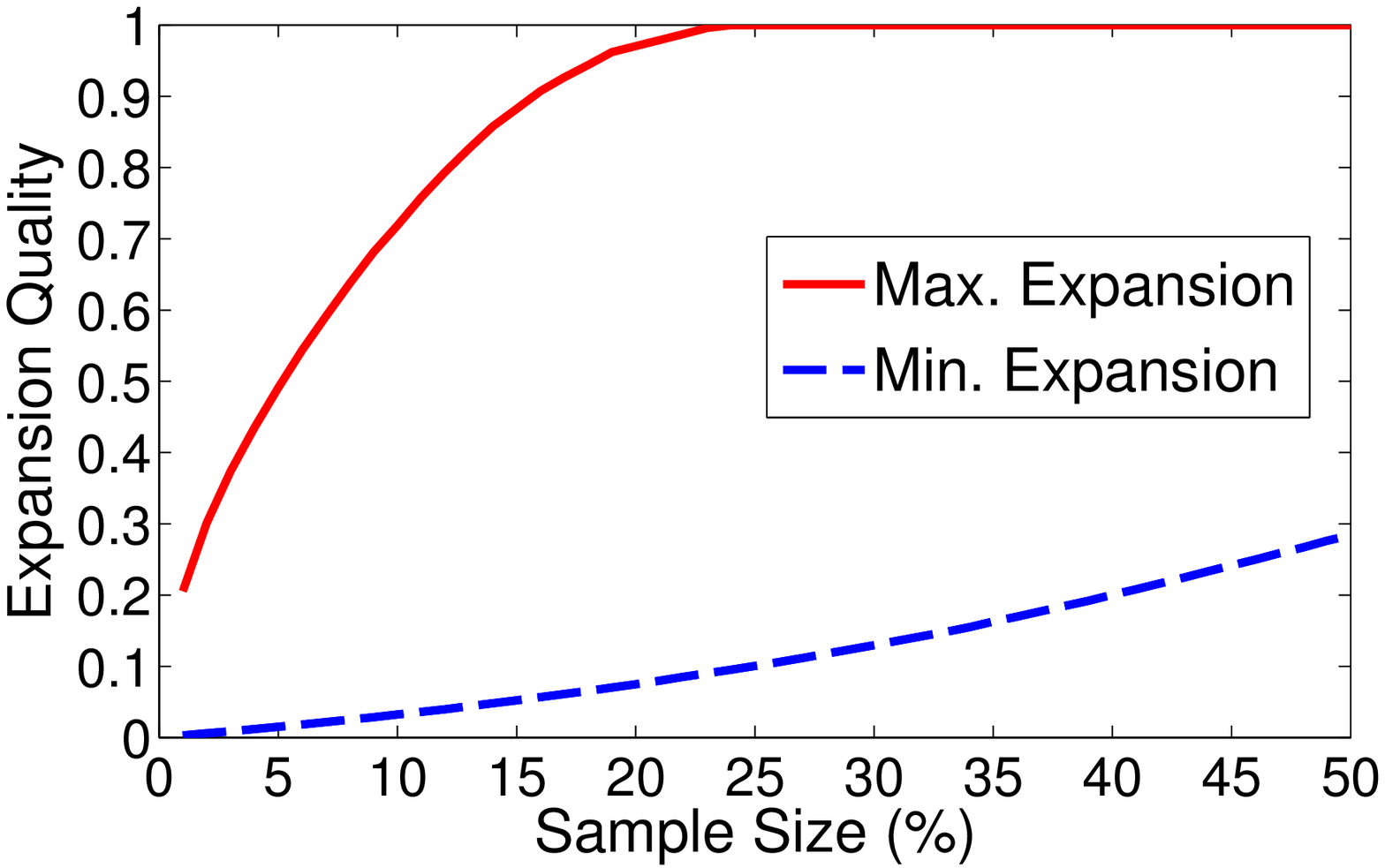}} \\ \vskip -0.01in
  \subfloat[Epinions]{\label{fig:esig.epinions}\includegraphics[width=0.2\textwidth]{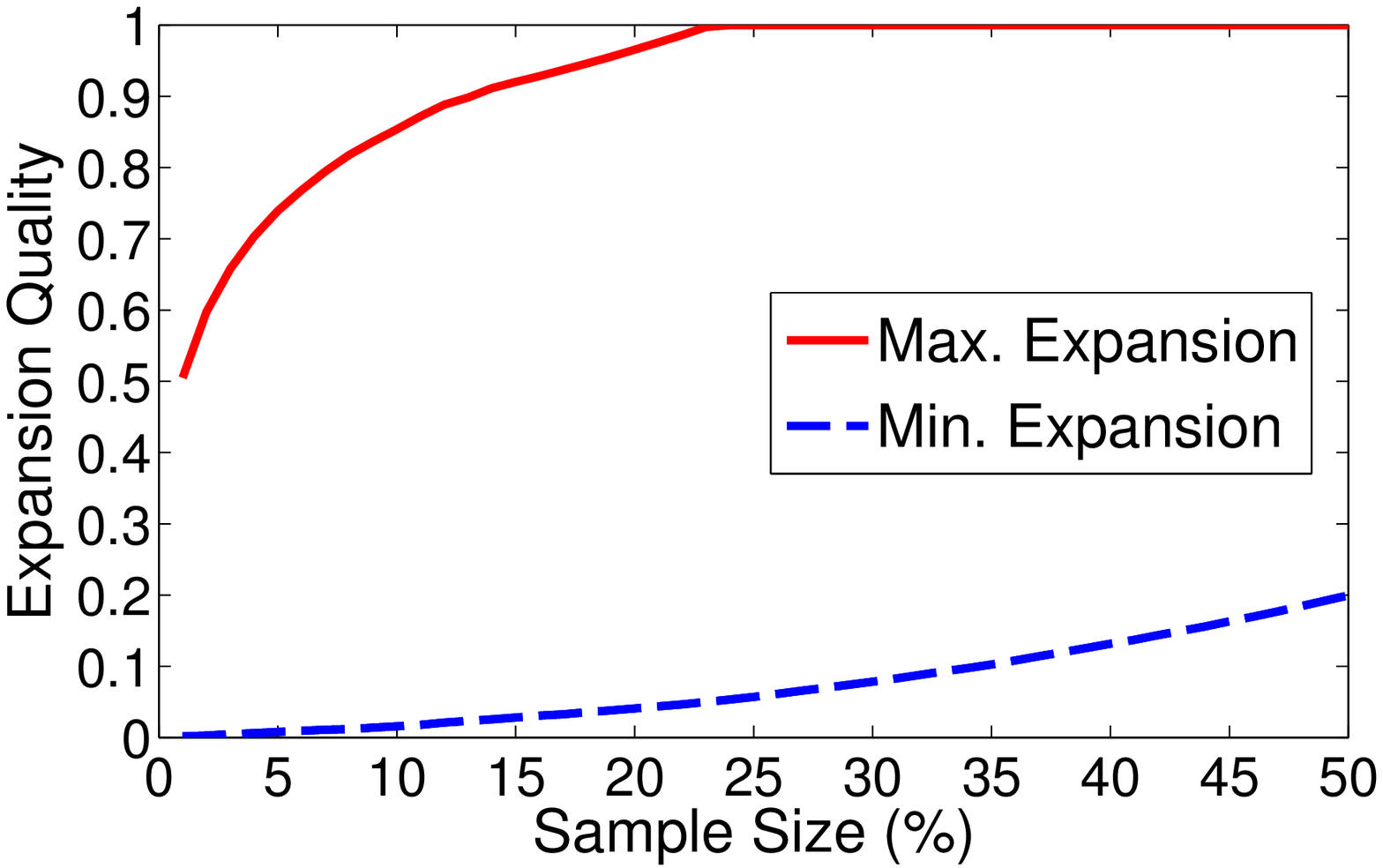}}\vspace{.05cm}
  \subfloat[Slashdot]{\label{fig:esig.slashdot}\includegraphics[width=0.2\textwidth]{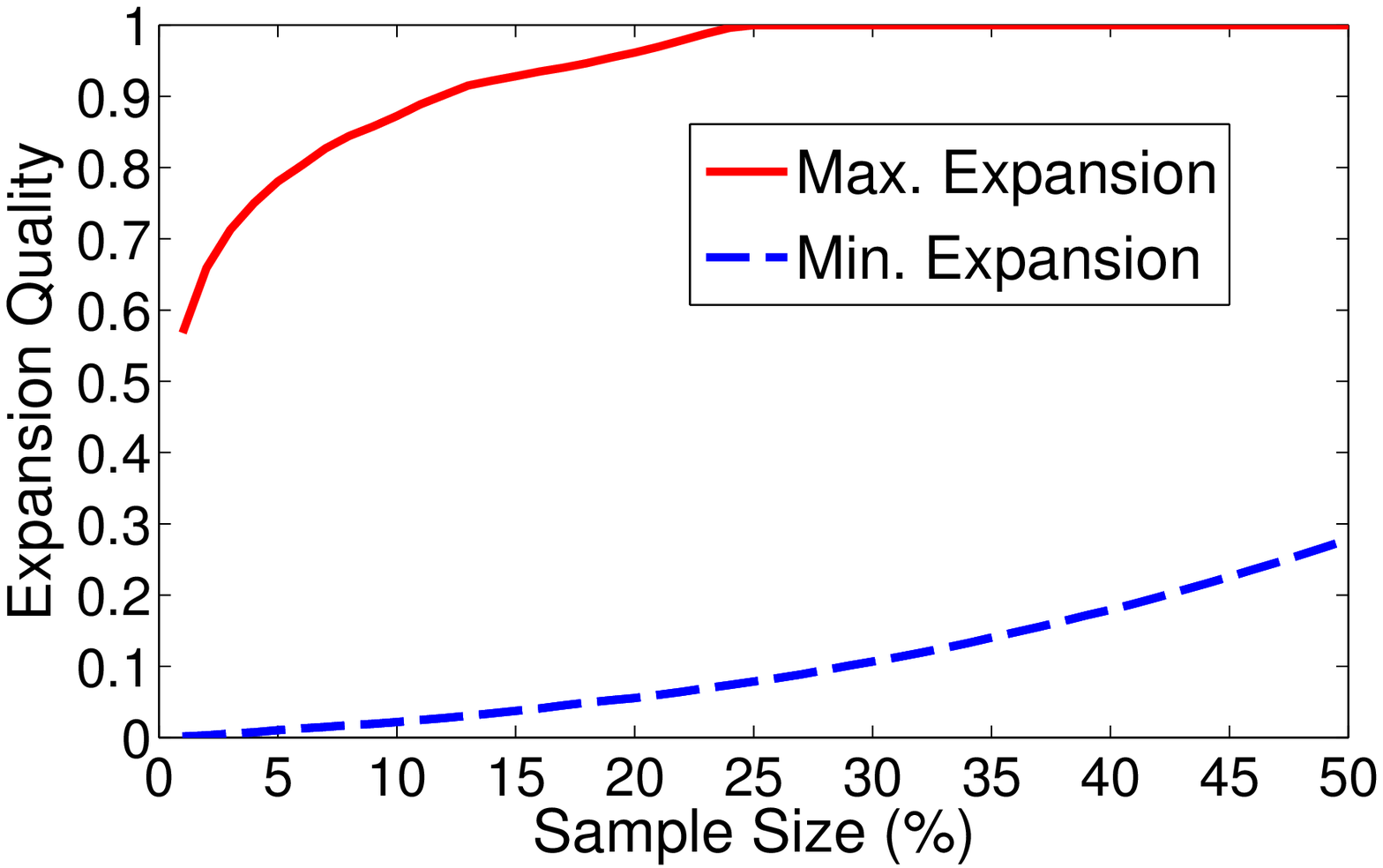}}
\caption{\emph{\esigcap s} for different networks.}
  \label{fig:esig.realnetworks}
\end{figure}

We now turn our attention to the \emph{\esig s} of real-world networks.  We examine eight different networks from diverse domains.  \emph{\esigcap s} for each network are shown in Figure \ref{fig:esig.realnetworks}.  We can immediately see that different types of networks exhibit very different expansion properties.  For instance, the size scale required to obtain a maximum \emph{expansion quality} of 1 in the Enron network is only $7\%$.  For the power grid, it is $49\%$.  

We also see that the \emph{minimum} expansion varies across networks.  We identify two different causes for low minimum expansion:  1) there is \emph{extreme} sparsity in the number of edges (imagine a simple line graph) or 2) the network is relatively sparse while exhibiting a high degree of clustering (imagine small sets of densely linked nodes linked together by \emph{sparse} connections).  Both cases result in a relatively low minimum expansion (as the neighborhood size ($|N(S)|$) will be small for most samples).  In the former case, the maximum expansion will tend to also be low (e.g. Power Grid).  In the latter case, we find the maximum expansion to be relatively higher (e.g. CondMat, Enron).  As will be discussed later, we posit the sparse links between relatively dense clusters in networks result in these higher values for maximum expansion.   Recall also that the minimum expansion is related to the classic definition of an expander graph:  every $S$ has expansion at least $\gamma$ in a $\gamma$-expander \cite{Hoory2006Expander}.  The co-authorship, email, and social networks, with higher clustering and consequent lower minimum expansion, do \emph{not}, then, appear to be classic expander graphs.  In Section \ref{sec:evaluation}, we will see whether or not this low minimum expansion affects searchability and information dissemination.

\section{Searching Networks}
\label{sec:searching}

We now address the problem of decentralized search in networks.  In a typical realistic scenario, starting from some initial node, we must locate some other node in the network \emph{without} full knowledge of global network topology.  Thus, we are unable to simply compute the shortest path, and we must hop from node to node until the destination node is found.  The running example application we employ is querying unstructured peer-to-peer file-sharing networks, where a search is comprised of sending a query message from node to node.  The destination node in question, for instance, might host a particular file of interest to the querier.  How can we locate this destination node?  Flooding the network with the query (where a node receiving a query forwards it to all neighbors) is provably unscalable and impractical \cite{Adamic2001Search,Schmid2007Structuring}.  In fact, when the music file-sharing service Napster became unavailable due to a court injunction in 2001, the Gnutella network (which employed a flooding-based search protocol at the time) crashed due to the large influx of former Napster users \cite{Schmid2007Structuring}. 

In Section \ref{sec:expansion}, through \emph{\esig s}, we have seen that it is often a relatively small set of nodes that is connected to a large portion of the network.  If one were able to easily locate this set of high expansion nodes, then the efficiency of search might vastly be improved (as this would quickly take us within one hop of many other nodes).  But, how can these nodes be accessed during the course of a decentralized search?  Our aforementioned greedy $(1-1/e)$-approximation algorithm to find high expansion nodes, shown in Algorithm \ref{alg:greedy}, assumes we have access to the network in its entirety (in which case decentralized search would not even be needed).  As mentioned, we are interested in cases where there is \emph{limited} knowledge of global network connectivity.  Therefore, we adapt the greedy $(1-1/e)$-approximation algorithm from Section \ref{sec:expansion} into a greedy search heuristic - one that does \emph{not} require full knowledge of network topology. We refer to this search heuristic as an \emph{expansion search}.   We compare the  \emph{expansion search} to several popular search strategies in complex networks.  These include a \emph{degree search} \cite{Adamic2001Search}, a breadth-first search (BFS) \cite{Jin2007Novel,Jiang2008LightFlood,Yang2002Improving}, and a random walk \cite{Adamic2001Search,Li2006Searching}.  We note that, in a BFS-based search strategy, there are \emph{multiple} copies of the search query traversing the network.  In contrast, for the remaining three search strategies, there is a \emph{single} copy of the search query.  For all search methods, unvisited nodes are always preferentially selected over previously visited nodes at each step in the search.  We now describe each search strategy in detail.  
~\\
\noindent
\textbf{Expansion Search (XS).}  In an \emph{expansion search}, the next node in the search is selected so as to maximize the expansion.  Let $S$ be the set of nodes visited thus far, let $N(S)$ be the neighborhood of the visited nodes, and let $c$ be the current, most recently visited node (where $c \in S$).  Then, in an \emph{expansion search}, the next hop is selected from among the unvisited neighbors of $c$ (i.e. $N(\{c\})-S$) as the node that maximizes the expansion.  That is, we visit node $v$ where $$v = \argmax_{v \in N(\{c\})-S} |N(\{v\}) - (N(S) \cup S)|$$  The key difference, then, is that the next hop is selected from the neighborhood of the current node $c$ (i.e. $N(\{c\})$), rather than all of $V-S$ (as is the case in the greedy approximation algorithm described in Algorithm \ref{alg:greedy}).  

~\\
\noindent
\textbf{Degree Search (DS).}  The degree-based search was proposed by Adamic et al. \cite{Adamic2001Search}.  At each step in the search, the search query is forwarded to the unvisited neighbor with the highest degree (i.e. largest number of neighbors).  That is, the next hop selected is node $v$ where $$v = \argmax_{v \in N(\{c\})-S} |N(\{v\})|.$$  Adamic et al. \cite{Adamic2001Search} analytically and empirically showed that, for power-law networks, if nodes with highest degree are preferentially selected during the search and visited first, substantial portions of the network can be covered and explored.  

~\\
\noindent
\textbf{Breadth-First Search (BFS).}  One type of search strategy used most often in practice is a breadth-first search \cite{Mitra2009Technological,Tsoumakos2006Analysis}.  In its simplest form, this involves flooding the network, where each node sends a copy of the query to each and every one of its neighbors.  These flooding and broadcast methods find targets quickly.  But, as we have already mentioned, they are highly unscalable due to the tremendous overhead incurred from redundant forwards (as each node forwards the query regardless of whether its neighbors have already received it).  As a result, a number of variations on flooding have been proposed to reduce this overhead (e.g. \cite{Jin2007Novel,Jiang2008LightFlood,Yang2002Improving}).  In this work, we evaluate a hypothetical BFS strategy in which there are \emph{no} redundant messages.  In other words, each message holder forwards a copy of the query only to those neighbors who have not yet received it, and all copies of the query terminate as soon as at least one copy of the query is successful and reaches its destination.  Note that this avoidance of redundant forwards and immediate termination are somewhat unrealistic for BFS or flooding strategies.  Unlike the other search methods we evaluate, there are \emph{multiple} copies of the query traversing the network in a BFS-based strategy.  And, with no information transfer between the various copies of the query, it is difficult to determine which neighbors have already seen the query or when one of the copies reaches the intended target.  Nevertheless, our implementation of pure BFS allows us to test the \emph{true} power of flooding-based strategies.  If this strategy, with its unrealistic and unfair advantage, still cannot match the performance of other search strategies, then BFS-based methods may not hold as much promise as previously thought, and their utility for exploring networks (e.g.  P2P, focused web crawling) should be possibly re-assessed.

~\\
\noindent
\textbf{Random Walk (RW).}  The final search strategy we evaluate is the random walk \cite{Adamic2001Search,Li2006Searching} in which the current node forwards the query to exactly one randomly selected neighbor.  We employ a \emph{self-avoiding} random walk \cite{Adamic2001Search} where the next  hop is selected randomly from among the neighbors who have not yet been visited in the search.  Note that, as opposed to BFS-based strategies, self-avoidance to eliminate redundant forwards is realistic here because there is a \emph{single} copy of the query traversing the network, within which a list of previously visited nodes can be stored.  (The same is true for self-avoidance in the \emph{expansion search} and the \emph{degree search}.)

~\\
We conclude this section with two final remarks.  First, for both the \emph{expansion search} and \emph{degree search}, each node must know both its neighbors \emph{and} its neighbors' neighbors.  This is required so that the \emph{expansion search} and \emph{degree search} can compute expansion and degree (respectively).  This, as it happens, is a modest and satisfiable requirement for many application domains.  For instance, in a P2P network, nodes must communicate with their neighbors when joining or leaving the network anyway and neighbor lists can be exchanged during this communication.    In fact, several existing search protocols exchange information with nodes at distances of even greater than two hops \cite{Adamic2001Search,Tsoumakos2006Analysis}.  Even in a social network, one typically is aware of friends of friends.

Second, as mentioned previously, unvisited nodes are always preferentially chosen over visited nodes in all four search strategies.  But, at some points during the search, it may be the case that all the neighbors of a given node are already visited.  There are several approaches to dealing with these situations.  The next step might be chosen uniformly at random from among the visited neighbors, for instance.  For the \emph{expansion search} and the \emph{degree search}, another approach is to select the ``best'' unvisited node from among the neighborhood of all previously visited nodes (i.e. if $S$ is the set of visited nodes, select a node $v \in N(S)$ with the highest degree or best expansion).  Note that, if using this approach, the partial topology of the network, learned during the course of the search, must be stored so that a path to the best next hop may be traversed.  During preliminary testing, we did not find a significant performance difference between the two.  Therefore, we only consider the former approach:  when all neighbors of a current node are visited, the next hop is selected uniformly at random from among these visited neighbors.

\section{Experimental Evaluation}
\label{sec:evaluation}

\subsection{Experimental Setup}
\label{sec:evaluation.setup}

We evaluate each search strategy on each network and track performance over time. Each node is assumed to know its neighbors and passes received messages to them based on one of the four search strategies.  The search ends when the message is passed to a neighbor of the target, at which time the message-holder can pass the message directly to its destination\footnote{In the context of P2P, we assume each node knows the \emph{identity} of its neighbors' neighbors, but not necessarily the \emph{files} stored by its neighbors' neighbors.}.  We track the cumulative nodes discovered\footnote{For the Experimental Evaluation section, we employ the normalized cumulative nodes discovered ($\frac{|N(S) \cup S|}{|V|}$) as the evaluation measure rather than the \emph{expansion quality} ($\frac{|N(S)|}{|V-S|}$).} at each step of a search, which is comprised of both the nodes visited and the neighbors of nodes visited.  We define a ``step'' in the search as a single hop taken by a single query message.  If there are multiple copies of the message (as in the case of a BFS or flooding strategy), then the number of steps is defined as the total number of hops taken by all copies of the message in the system. Note that this setup is somewhat of a worst case scenario, as we are assuming there is but a single node in the entire network capable of satisfying a given search query.  In the context of a P2P network, for instance, we are assuming that there is a single file residing on a single node in the whole network that must be located.  As a result, actual performance in real applications, where multiple nodes can satisfy a search query, will be much higher.  The extent will be domain-specific and depend on the extent of object (or file) replication in the network.  This setup, then, allows us to evaluate the performance of each search strategy \emph{independent} of the effects of extraneous factors such as replication.

\subsection{Experimental Results}
\label{sec:evaluation.results}
\subsubsection{On the Performance of Search Strategies}
\label{sec:evaluation.results.performance}

We first examine the relative performance of each search strategy on each network.  Table \ref{tab:searchresults} shows the number of steps required to discover $20\%$, $35\%$, and $50\%$ of the nodes in the network.  As can be seen, the \emph{expansion search} (XS) exhibits the best overall performance.  We also find a clear performance difference between the conventional search strategies (BFS and RW) and the less conventional approaches (XS and DS).  We discuss each separately.

~\\
\textbf{XS and DS Performance}\\
Overall, we find the XS and DS strategies to exhibit the best general performance with the XS approach faring better.  On most of the networks, the XS strategy either exceeds or ties the performance of other approaches. This leads us to a natural question: what causes performance differences between XS and DS?  On networks in which XS and DS perform similarly, high degree nodes will tend to link to different sets of nodes (in which case high degree nodes and high expansion nodes will tend to be one and the same and will discover a similar amount of nodes).  On the other hand, for networks where XS exceeds the performance of DS, we posit that these nodes may be more likely to have similar neighbors, in which case a high degree node may, in fact, have \emph{low expansion} if it links to neighbors already seen during the search.  In these cases, the XS strategy will discover more nodes.  Overall, despite the modestly better performance of the XS method, we find the DS strategy performs exceedingly well, which indicates that the former case may be more common in real-world networks.  That is, on real-world networks, a \emph{degree search} may do well in finding high expansion nodes without explicitly looking.  

The one network on which neither XS nor DS performs the best is the power grid.  The power grid seems to be the least well-connected network evaluated (with mean degree of only $2.7$).  In fact, it has such low connectivity that only a systematic BFS does best in exploring the network.  

~\\
\textbf{BFS and RW Performance}\\
Conventional approaches to searching networks such as P2P systems include those based on random walks (RW) and breadth-first search (BFS) \cite{Mitra2009Technological}.  As mentioned, flooding strategies, based on BFS, are used most often in real applications, as they tend to find answers quickly.  BFS is also pervasively used in web crawling and graph sampling.  It is striking, then, that our idealized version of BFS, one that avoids redundant communications and immediately terminates upon success, still cannot outperform other approaches (on all but the power grid).  In general, we find that the BFS and RW approaches exhibit a relatively lower \emph{expansion quality} as compared to XS, fail to explore the network as well as other strategies in the same number of message forwards, and, consequently, discover less nodes.

\begin{figure}[htb]
  \centering
  \subfloat[Erdos-Renyi] {\label{fig:xplr.er}\includegraphics[width=0.2\textwidth]{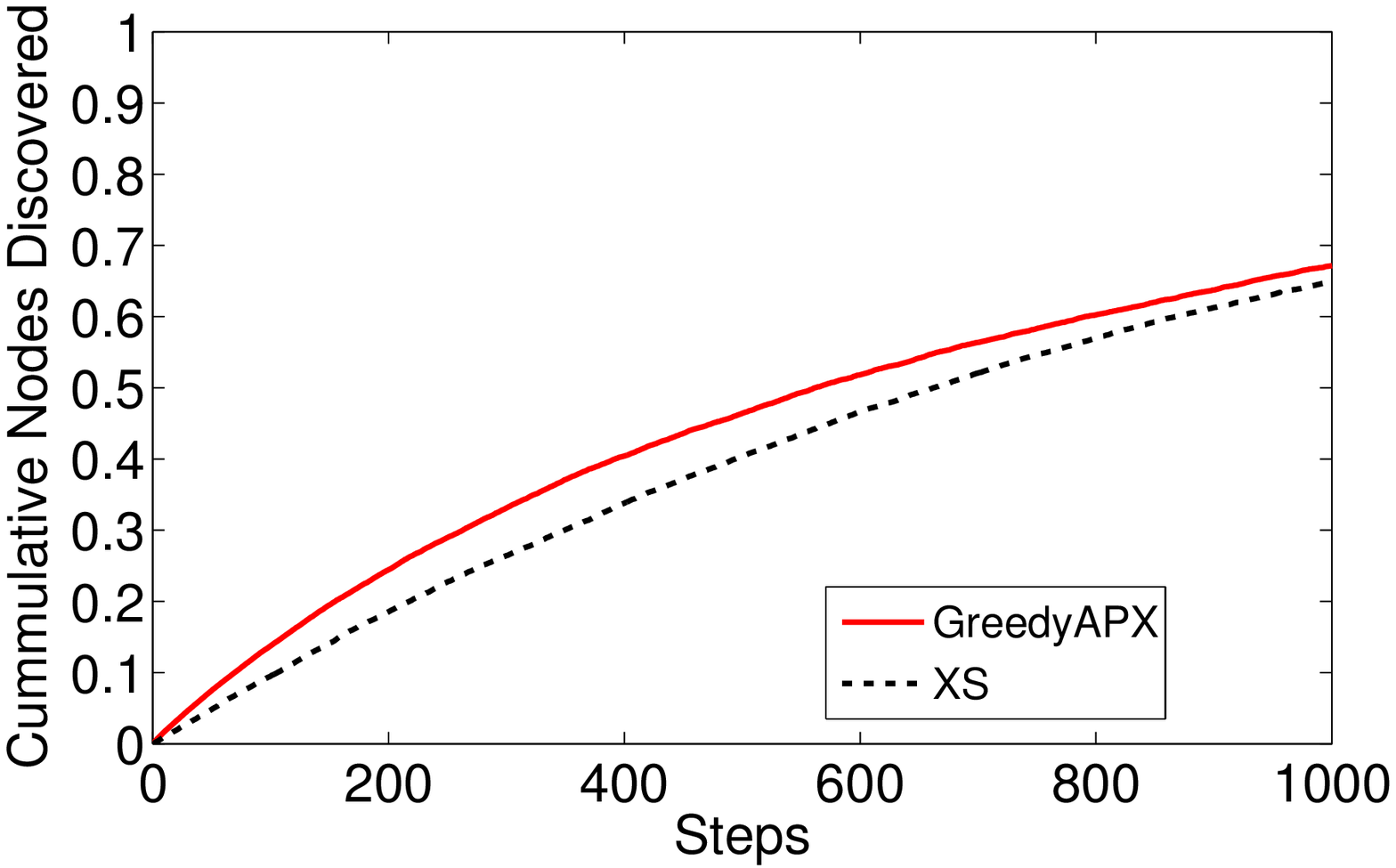}} \vspace{.05cm}
  \subfloat[Barabasi-Albert]{\label{fig:xplr.ba}\includegraphics[width=0.2\textwidth]{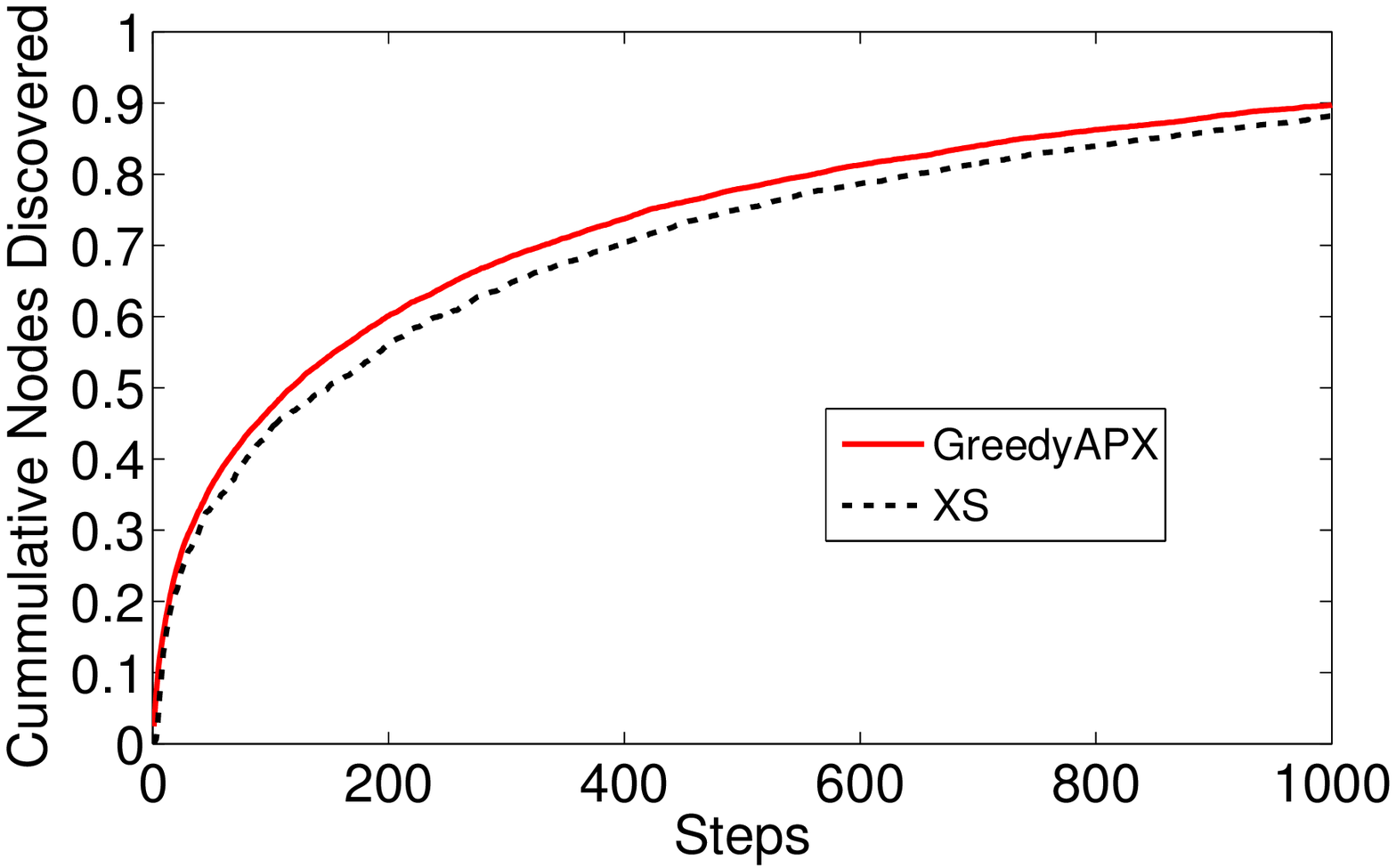}} \\ \vskip -0.01in 
  \subfloat[C. elegans] {\label{fig:xpor.celegans}\includegraphics[width=0.2\textwidth]{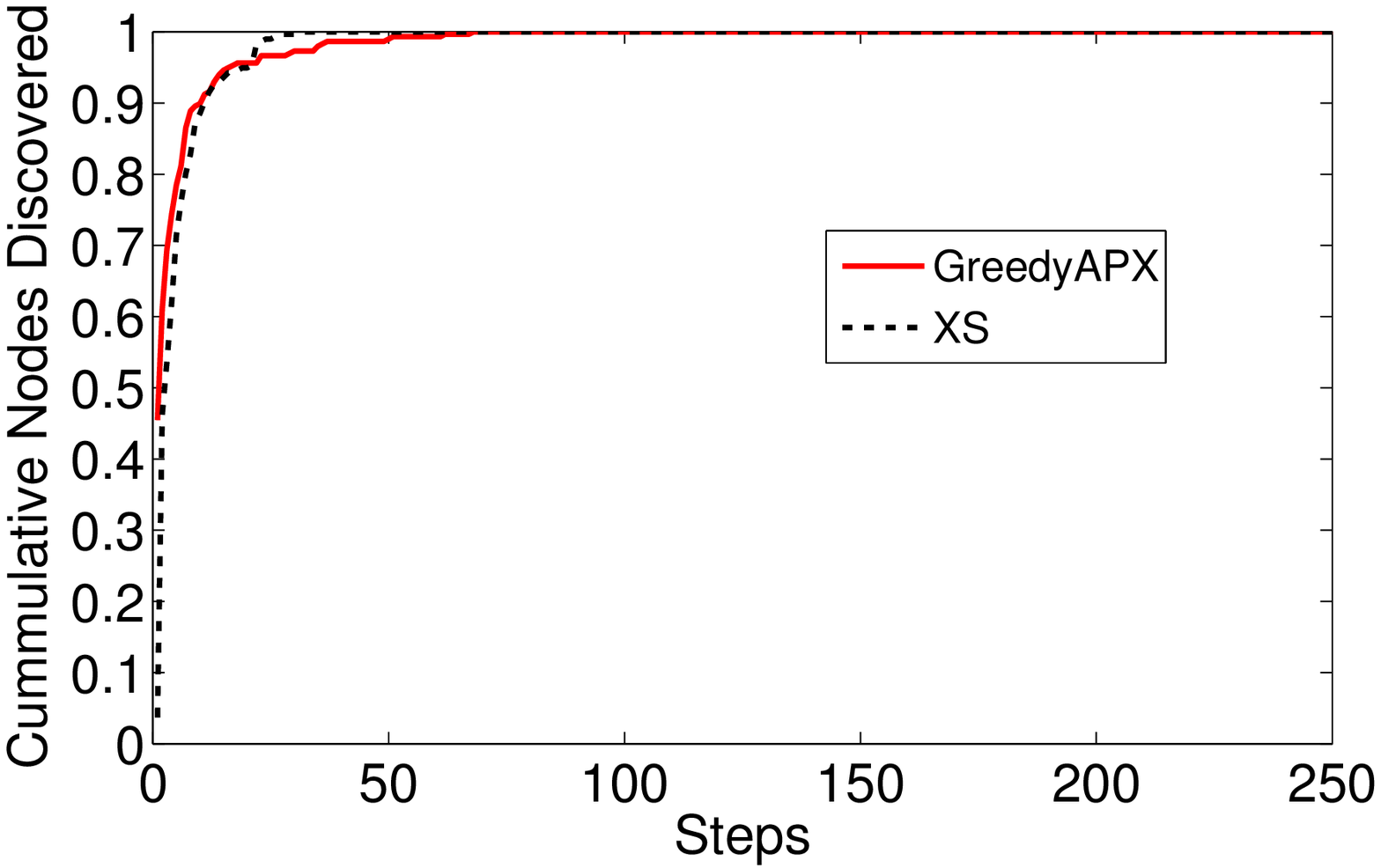}} \vspace{.05cm}
  \subfloat[Power Grid]{\label{fig:xpor.power}\includegraphics[width=0.2\textwidth]{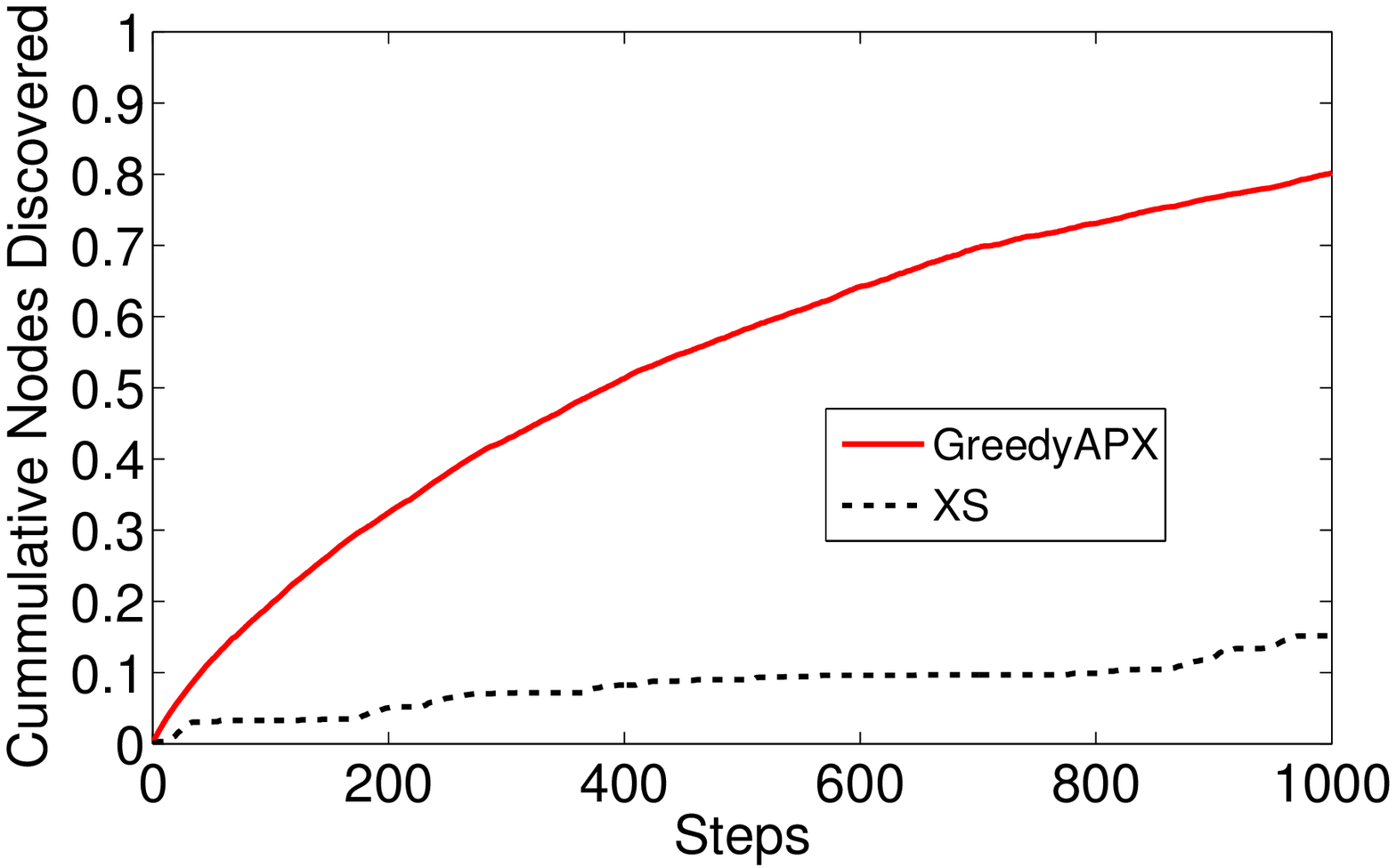}} \\ \vskip -0.01in 
  \subfloat[CondMat] {\label{fig:xplr.condmat}\includegraphics[width=0.2\textwidth]{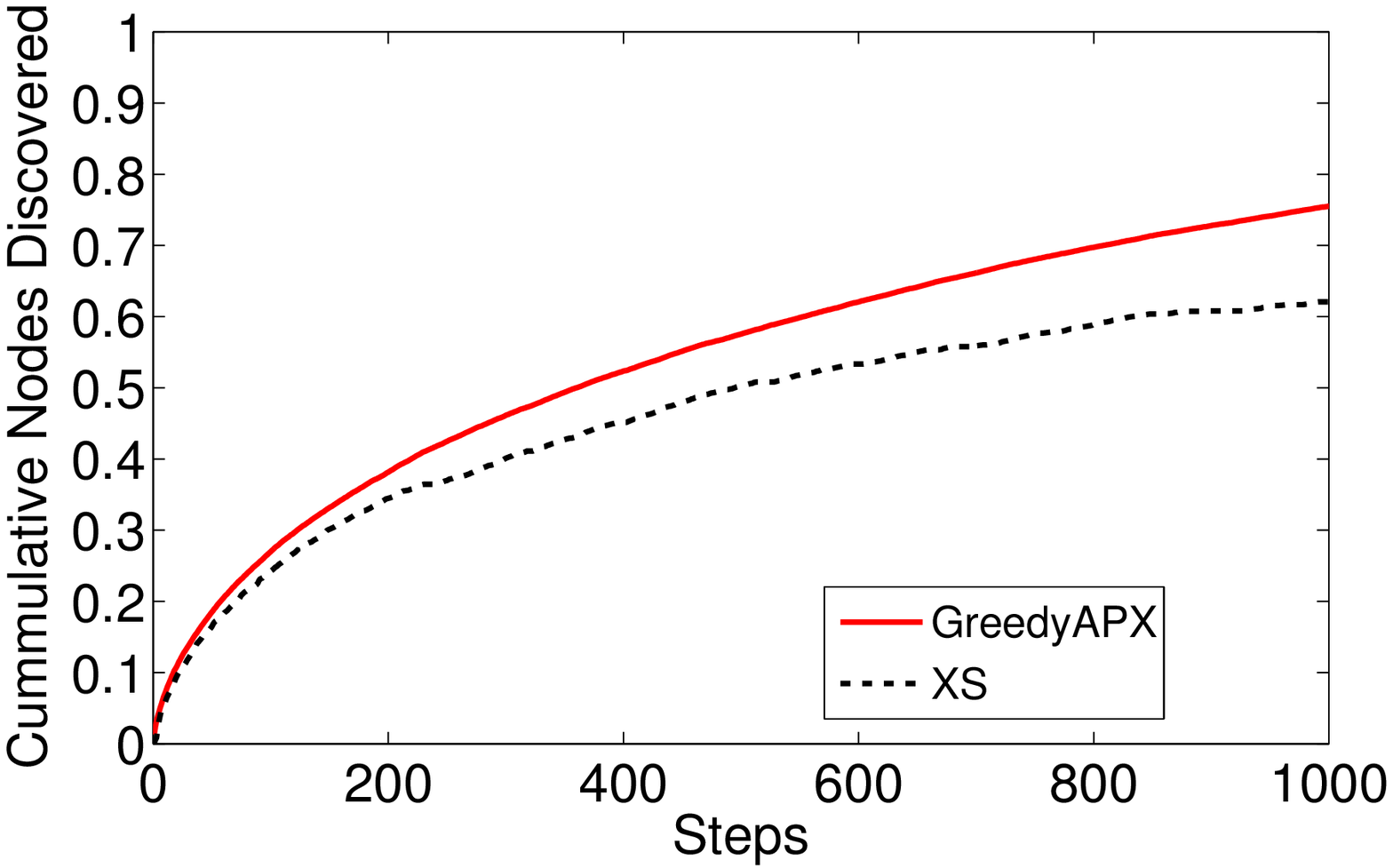}} \vspace{.05cm}
  \subfloat[Enron] {\label{fig:xplr.enron}\includegraphics[width=0.2\textwidth]{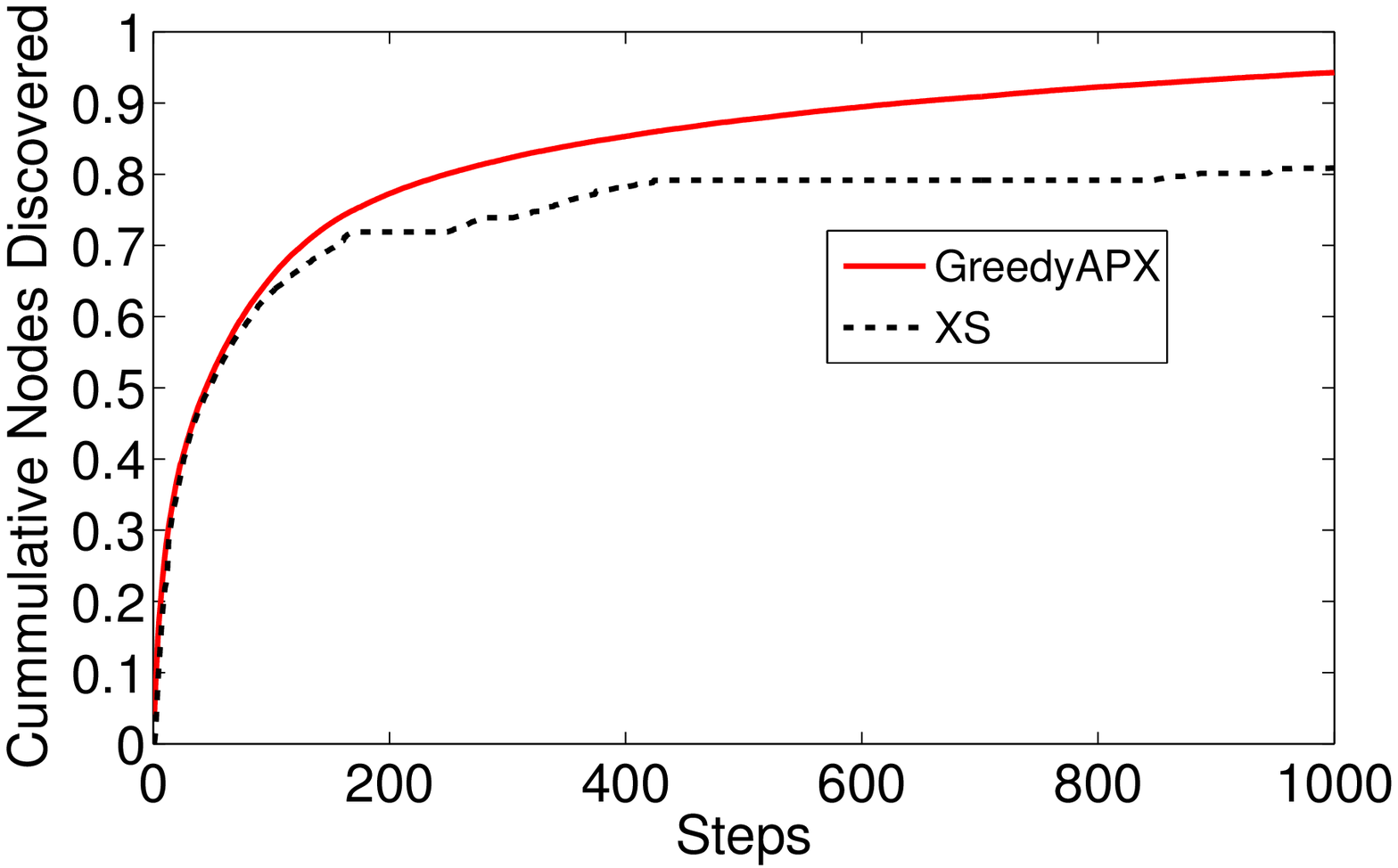}}\\ \vskip -0.01in
  \subfloat[HEPPh]{\label{fig:xplr.hepph}\includegraphics[width=0.2\textwidth]{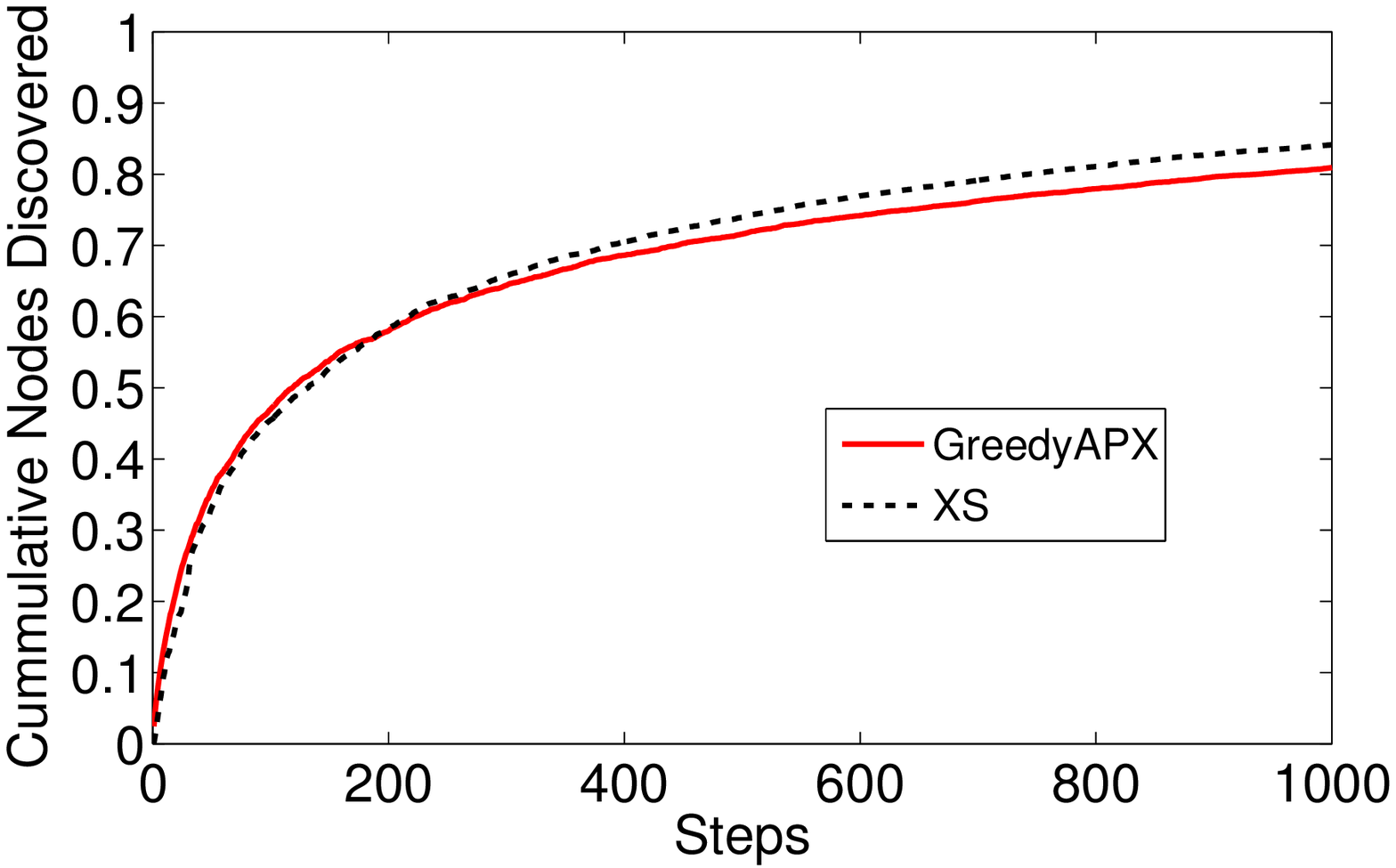}} \vspace{.05cm}
  \subfloat[Gnutella]{\label{fig:xplr.gnutella31}\includegraphics[width=0.2\textwidth]{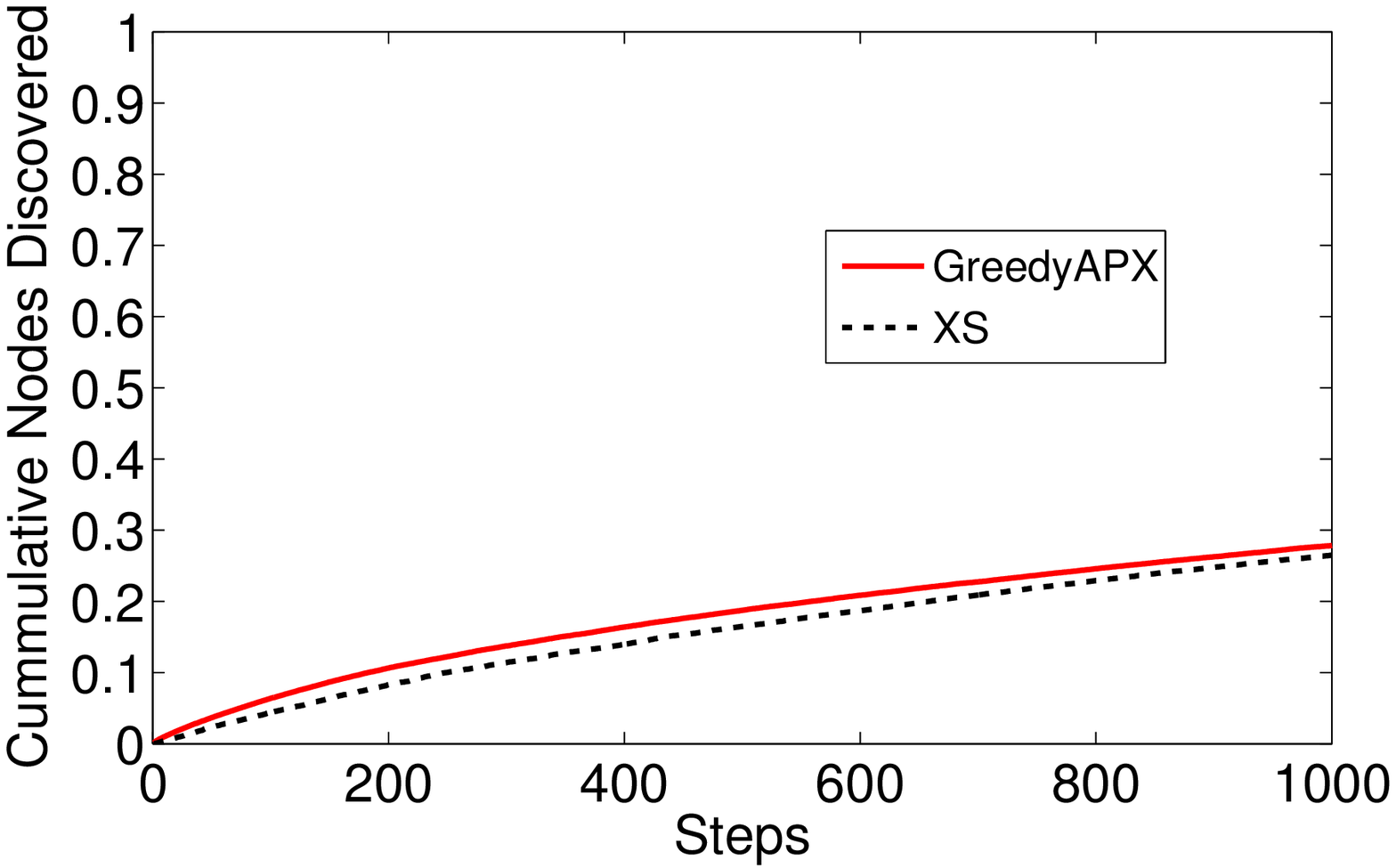}}\\ \vskip -0.01in
  \subfloat[Epinions]{\label{fig:xplr.epinions}\includegraphics[width=0.2\textwidth]{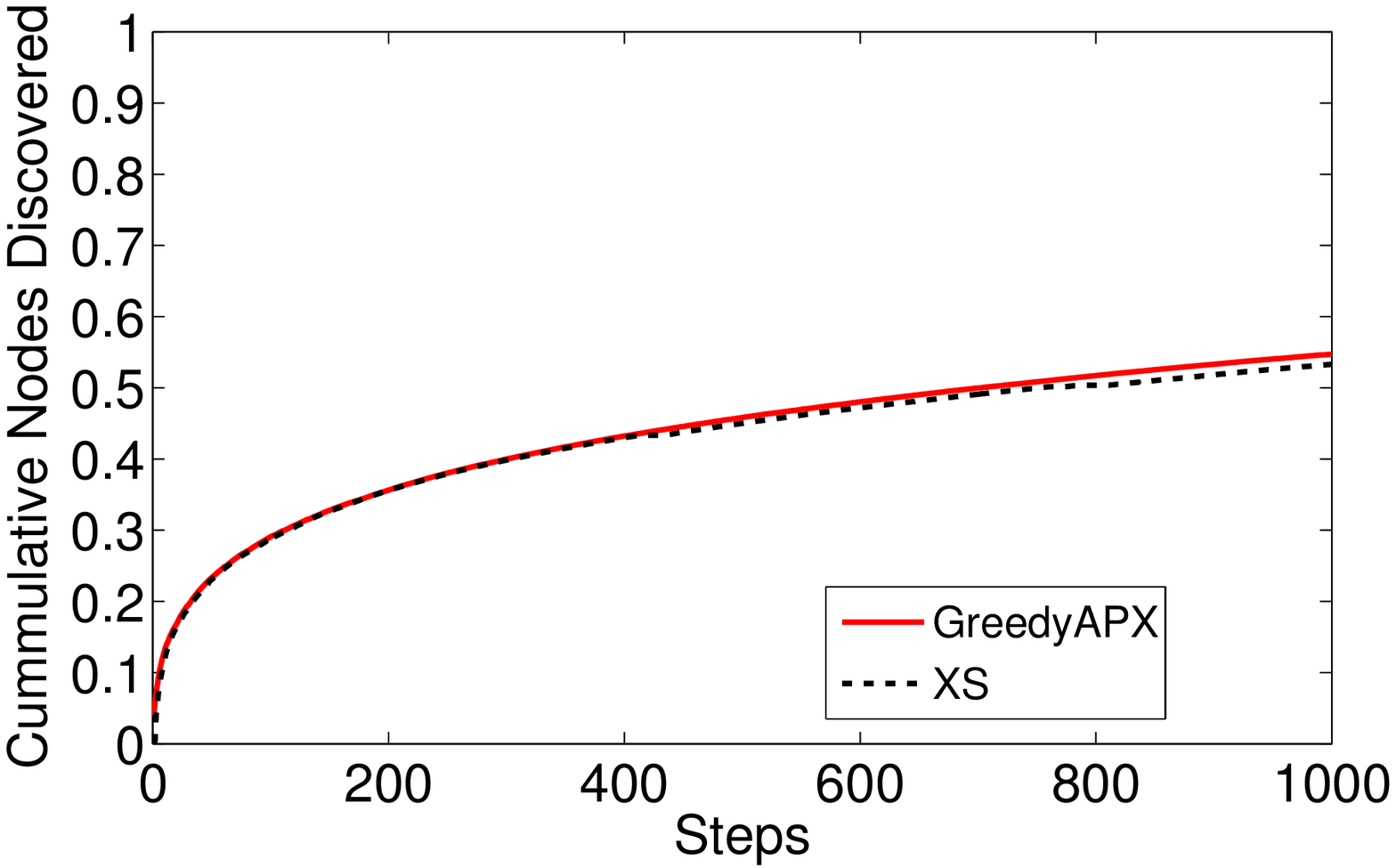}}\vspace{.05cm}
  \subfloat[Slashdot]{\label{fig:xplr.slashdot}\includegraphics[width=0.2\textwidth]{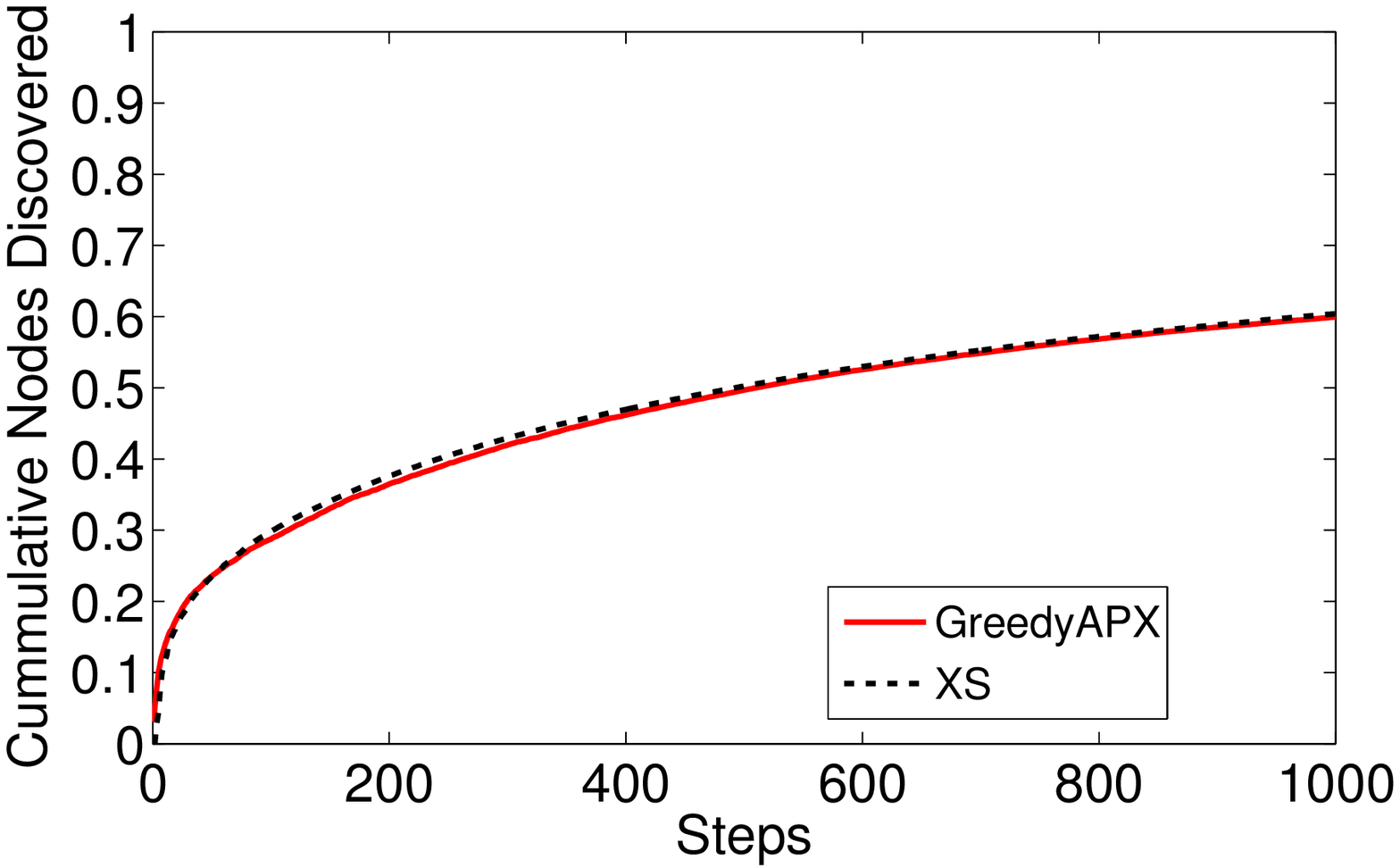}}
\caption{[\textbf{Best viewed in color.}]  Comparison of {\sc \greedy} and XS for first 1000 steps of a search.  In most cases (save for the power grid), XS strategy closely matches {\sc \greedy} (our best approximation for the maximum expansion).}.
  \label{fig:xpl}
\vskip -0.15in
\end{figure}

~\\
\textbf{Comparison to \greedy}\\
Given the relatively better performance of the XS strategy as compared to other methods, we now examine the extent to which it matches the performance of {\sc \greedy} (our best known approximation of the maximum expansion in a network).  Figure \ref{fig:xpl} shows the cumulative nodes discovered by {\sc \greedy} and XS for the first $1000$ steps.  Interestingly, the XS strategy, which hops from node to node and performs the search \emph{without} complete access to the network in its entirety, often comes close to the performance of {\sc \greedy} (which \emph{does} have random access to the entire network).  Once again, the most salient exception is the power grid, which seems to be the least searchable network evaluated.  We discuss network searchability next.

\subsubsection{On the Searchability of Networks}
\label{sec:evaluation.results.searchability}

~\\
\textbf{Expansion and Searchability}
\begin{sloppypar}
From Figures \ref{fig:esig.randomgraphs} and \ref{fig:esig.realnetworks} and Table \ref{tab:searchresults}, we can see that  the magnitude of maximum expansion (as approximated by {\sc \greedy}), corresponds remarkably well to the extent to which each network is searchable.  On any given network, when the maximum expansion is low, \emph{all} search strategies perform significantly worse.  On the other hand, when the maximum expansion is high, \emph{all} search strategies fare relatively better.  The \emph{\esig s}, then, correctly infer the ease of search and information dissemination in a network.  
\end{sloppypar}
It is also striking to find that it is the \emph{maximum} expansion (rather than the \emph{minimum} expansion) most responsible for the level of searchability in a network.  The classic definition of an expander graph is based on \emph{minimum expansion}.  Recall that a graph is a $\gamma$-expander if $|N(S)| \geq \gamma |S|$ for each $S \subset V$ where $|S| \leq \frac{|V|}{2}$ \cite{Hoory2006Expander}. In the literature, expander graphs and minimum expansion are often connected to the ease of dissemination in network (e.g. \cite{Chierichetti2010Rumour,Barret2007Fighting}).  For instance, \cite{Barret2007Fighting} has claimed social networks to be expander graphs as a means to explain the ease of diffusion across them.  In contrast, our work shows that social networks are \emph{not} classic expander graphs and have a low \emph{minimum} expansion due to clustering.  Moreover, we find that it is the \emph{maximum} expansion, not the minimum expansion, that is related to efficient searchability in social networks and other graphs.  

\begin{figure*}[htb]
  \centering
  \subfloat[Expansion Search (XS)] {\label{fig:trace.xs}\includegraphics[width=0.3\textwidth]{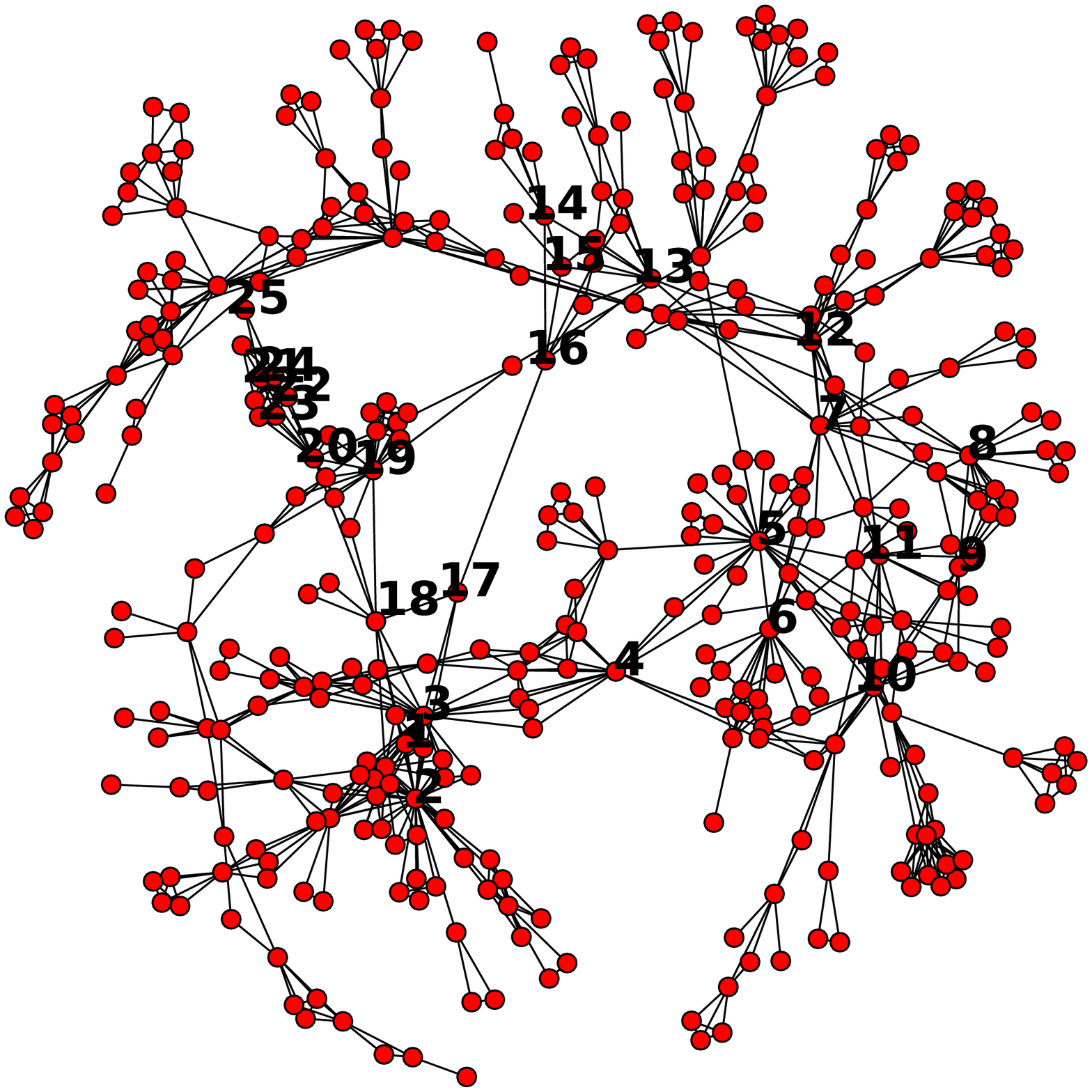}} ~~~~~~~~~~
  \subfloat[Random Walk (RW)]{\label{fig:trace.rw}\includegraphics[width=0.3\textwidth]{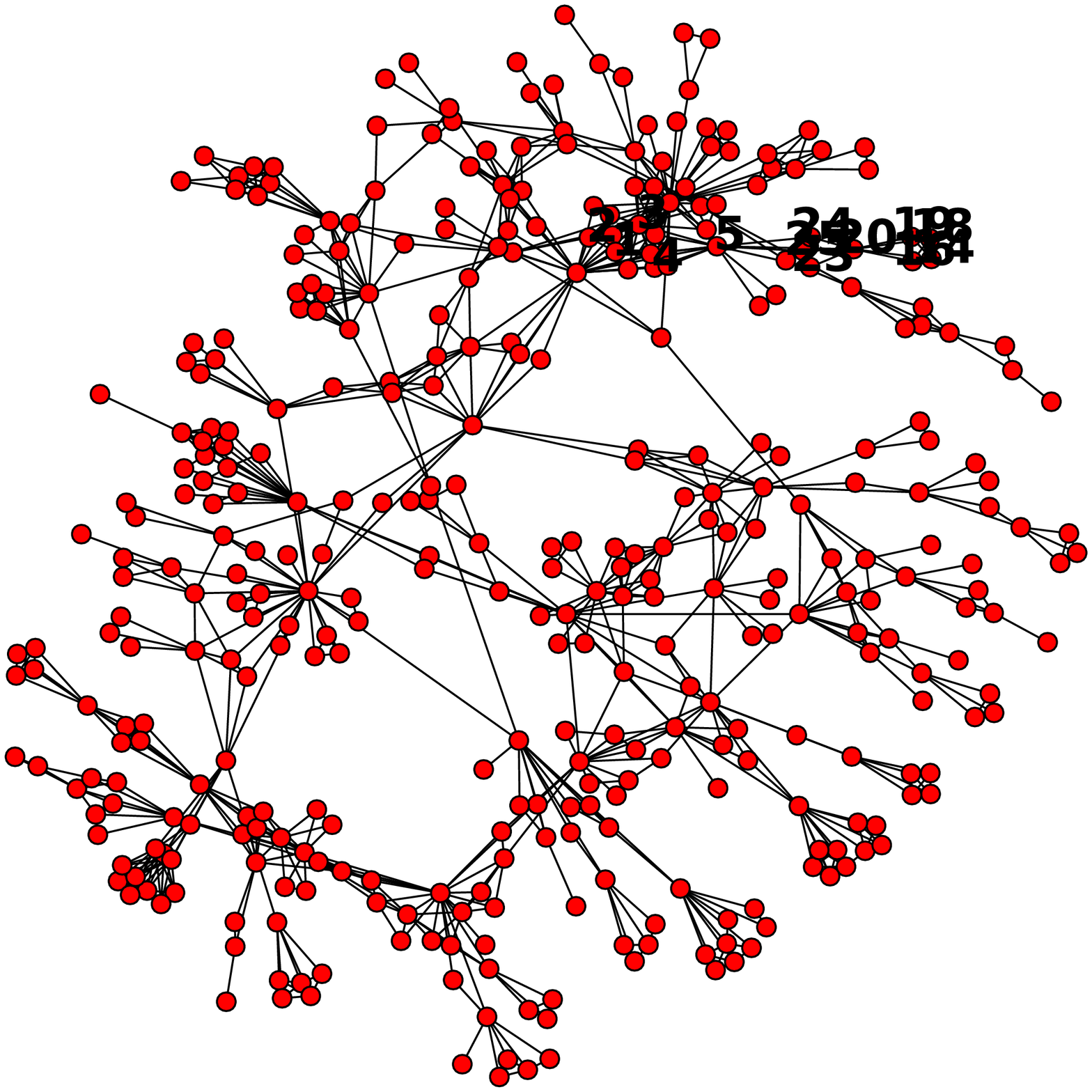}} 
\caption{Numbers on each plot show the trace of the first 25 steps in a search by an \emph{expansion search} (XS) and a self-avoiding random walk (RW).  Both searches were started from the same initial source node.  The XS strategy explores a wider portion of the network and more clusters in the same number of steps}.
  \label{fig:trace}
\vskip -0.15in
\end{figure*}

~\\
\textbf{Structural Properties and Possible Explanations}
\\
It is both surprising and ironic that the Gnutella network, which exists for the very purpose of search, turns out to be one of the \emph{least} searchable networks we evaluated.  The only other network exhibiting \emph{less} searchability is the power grid.  At the other end of the spectrum, the C. elegans and Enron networks appear to be the \emph{most} searchable.  As shown in Table \ref{tab:searchresults}, for both Enron and C. elegans, all four search strategies are able to discover half the network in a very small number of steps.  What causes a network to exhibit high maximum expansion and good searchability? 

One of the more obvious explanations is that denser, more well-connected networks tend to be more searchable than extremely sparse networks.  Nodes have larger neighborhoods in denser networks and are, therefore, easier to explore.  This is true for the same reason a clique is intuitively more searchable than a long sequence or chain of nodes each connected by a single edge. For instance, the ``unsearchable'' power grid has a density of $0.0005$ and mean degree of only $2.7$ whereas the C. elegans network has a density of $0.05$ and mean degree of $14.5$. The Gnutella network, like the power grid, is also relatively more sparse than other networks of equivalent size.

Density, however, fails to explain the whole story.  Consider the ER and BA graphs.  Both were constructed to have similar densities but exhibit different \emph{\esig s} (see Figure \ref{fig:esig.randomgraphs}) and, correspondingly, different degrees of searchability (see Table \ref{tab:searchresults}).  By virtue of its skewed degree distribution, the BA model seems to exhibit better searchability than that of the ER model, and the effect of degree distributions on search and dissemination is well-known (e.g. \cite{Barabasi1999Emergence,Adamic2001Search}).  By visiting well-connected hubs, one can quickly cover significant portions of a network.  But, once again, degree distributions fall short in adequately explaining the ease of search.  Many of the networks considered exhibit skewed, heavy-tailed degree distributions (e.g. Enron, Epinions), but, nonetheless, exhibit different levels of searchability.  Surprisingly, in stark contrast to previously held beliefs (e.g. \cite{Hui2006Smallworld}), even average path length fails to fully explain searchability.  A number of networks have very similar average path lengths (see Table \ref{tab:datasets}), but very different levels of searchability (see Table \ref{tab:searchresults}). 

Unlike density and degree distributions, the effect of clustering on searchability and dissemination is less studied and more nebulous.  As mentioned in Section \ref{sec:evaluation.results.performance}, based on our results, we reason that clustering can also facilitate searchability.  Real-world networks often exhibit what is known as \emph{community structure} \cite{Wasserman2005Models,Girvan2002Community}.  Intuitively, a community in a network is a cluster of nodes more densely connected to each other than other nodes and exhibit higher clustering coefficients than one would expect at random \cite{Wasserman2005Models,Girvan2002Community}.  By this intuitive definition, nodes in the same community will be expected to share more neighbors than nodes in different communities (by virtue of the dense connections within clusters and lower conductance).  As a result, if one were to visit a small number of nodes from many different communities, the expansion (and, therefore, conductance) of these visited nodes would be high and many nodes would be discovered in the search.  By searching based on expansion, more communities (and, consequently, larger portions of the network) are explored\footnote{This relationship between the maximum expansion and community structure has been demonstrated in \cite{Maiya2010Sampling}.}, and this can be demonstrated.  Consider the network theory co-authorship network \cite{Newman2006Finding}, a small, sparse network considered by many to exhibit some degree of community structure. Figure \ref{fig:trace}  shows a typical path taken by both an \emph{expansion search} (XS) and a random walk (RW) on this network.  The XS strategy, by attempting to maximize expansion, jumps across the boundaries between different clusters more easily and is able to explore larger portions of the network.  In this way, clustering and community structure, like high density and skewed degree distributions, can facilitate searchability in a network.  However, the only common thread and unifying theme that fully and consistently explains searchability across different networks is the singular concept of \emph{expansion}.

\begin{table*}[htb]
\centering
\begin{tabular}{l|cccc|cccc|cccc} \hline \hline
{\bf ~} & \multicolumn{4}{c|}{\bf 20\%} & \multicolumn{4}{c|}{\bf 35\%} & \multicolumn{4}{c}{\bf 50\%}  \\
        & XS & DS & RW & BFS & XS & DS & RW & BFS & XS & DS & RW & BFS \\ \hline 
        & ~ & ~ & ~ & ~ & ~ & ~ & ~ & ~ & ~ & ~ & ~ & ~ \\  
   ER          & \textbf{218} & 224 & 366 & 386  & \textbf{417} & 443 & 711 & 738 & \textbf{662} & 720 & 1141 & 1188 \\
   BA          & \textbf{18} & \textbf{18} & 154 & 91  & 59 & \textbf{56} & 288 & 285 & 149 & \textbf{145} & 537 & 393 \\
   C. eleg.    & \textbf{2} & \textbf{2} & \textbf{2} & 3  & \textbf{2} & \textbf{2} & 4 & 7 & \textbf{3} & \textbf{3} & 9 & 8 \\
   Power       & 1394 & 1450 & 1271 & \textbf{649}  & 2220 & 2370 & 2794 & \textbf{1332} & 8051 & 6151 & 5148 & \textbf{2091} \\
   CondMat     & \textbf{72} & 93 & 474 & 413  & \textbf{208} & 317 & 1064 & 1071 & \textbf{495} & 827 & 2248 & 2336 \\
   Enron       & \textbf{9}   & 10 & 125 & 266  & \textbf{20} & 22 & 342 & 801 & \textbf{49} & 58 & 559& 1941 \\
   HEPPh       & \textbf{26} & 37 & 204 & 446   & \textbf{56}  & 80 & 469 & 1366 & \textbf{132} & 250 & 825 & 2205 \\
   Gnutella    & \textbf{659} & 720 & 1788 & 1730  & \textbf{1577} & 1836 & 3897 & 3829 & \textbf{2930} & 3615 & 7191 & 6875 \\
   Epinions    & \textbf{34}  & 48  & 281  & 590 & \textbf{189} & 344 & 1059 & 2679 & \textbf{752} & 1213& 3029 & 5948 \\
   Slashdot    & \textbf{32}  & 43  &  241 & 338  & \textbf{163} & 239 & 859 & 1612 & \textbf{492} & 725 & 1997 & 4210 \\ \hline \hline
 \end{tabular}
 \caption{Number of steps to discover 20\%, 35\%, and 50\% of the network.  The best (i.e. lowest) value is in highlighted for each dataset.  Overall, XS performs best.  The variance for XS and DS was significantly small and standard error is omitted for ease of illustration.  (Standard error for RW/BFS was larger, but not so large that either became a candidate for the best or even second-best performer.)}
 \label{tab:searchresults}
\end{table*}

\section{Conclusions}
We have introduced the concept of \emph{\esig s} and have used them to study the effect of expansion on decentralized search in networks.  We have shown that it is the magnitude of maximum expansion (rather than minimum expansion) that corresponds to the extent to which a network is efficiently searchable.  Moreover, we have shown that traditional graph properties such as average path length and skewed degree distributions fail, by themselves, to fully explain the level of searchability in a network.  Finally, we have shown that a search strategy based on maximizing expansion covers the network far better than some typical approaches to decentralized search.  For future work, we plan to further investigate the interplay between expansion and various graph-theoretic properties and their effect on dissemination.  


\balance
\bibliographystyle{abbrv}

\end{document}